\documentclass[aps, preprint, nofootinbib,preprintnumbers,eqsecnum]{revtex4}

\usepackage{color}

\usepackage[
      colorlinks=true,
      linkcolor=blue,
      urlcolor=blue,
      filecolor=black,
      citecolor=red,
      pdfstartview=FitV,
      pdftitle={},
        pdfauthor={Thomas Faulkner, Nabil Iqbal},
        pdfsubject={},
        pdfkeywords={},
        pdfpagemode=None,
        bookmarksopen=true
      ]{hyperref}

\usepackage[normalem]{ulem}
\usepackage{amsmath}
\usepackage{enumerate}
\usepackage{amsfonts}
\usepackage{yfonts}

\usepackage{subfigure}
\usepackage{psfrag}

\usepackage{epsfig}
\usepackage[latin1]{inputenc}
\usepackage{float}
\usepackage{graphicx}
\usepackage{cancel}
\usepackage{mathrsfs}
\usepackage{amssymb}
\usepackage{amsfonts}
\usepackage{amsmath}
\usepackage{slashed}

\usepackage{graphicx}
\usepackage{bm}

\def\({\left(}
\def\){\right)}
\def\[{\left[}
\def\]{\right]}
\def\<{\langle}
\def\>{\rangle}

\newcommand\half{{\ensuremath{\frac{1}{2}}}}
\newcommand\p{\ensuremath{\partial}}

\newcommand{\be}{\begin{equation}}
\newcommand{\ee}{\end{equation}}
\newcommand{\bea}{\begin{eqnarray}}
\newcommand{\eea}{\end{eqnarray}}
\newcommand{\bwt}{\begin{widetext}}
\newcommand{\ewt}{\end{widetext}}

\newcommand{\bi}{\begin{itemize}}
\newcommand{\ei}{\end{itemize}}
\newcommand{\ben}{\begin{enumerate}}
\newcommand{\een}{\end{enumerate}}
\newcommand{\bca}{\begin{cases}}
\newcommand{\eca}{\end{cases}}
\newcommand{\bln}{\begin{align}}
\newcommand{\eln}{\end{align}}
\newcommand{\bst}{\begin{split}}
\newcommand{\est}{\end{split}}

\newcommand\ep{\epsilon}

\newcommand\lam{\lambda}
\newcommand\Lam{\Lambda}
\newcommand\om{\omega}

\newcommand\ga{{\ensuremath{{\gamma}}}}

\newcommand\Th{{\Theta}}
\def\th{{\theta}}

\newcommand\ha{{\half}}

\def\le{\left}
\def\ri{\right}

\newcommand\sD{{\ensuremath{{\mathcal D}}}}

\newcommand\sG{{\ensuremath{{\mathcal G}}}}

\newcommand\sO{{\ensuremath{{\mathcal O}}}}

\newcommand{\ka}{{\kappa}}

\newcommand{\x}{{x}}

\def\htt{{h^\tau_{\,\tau}}}
\def\hxx{{h^\x_{\,\x}}}
\def\hxt{{h^\x_{\,\tau}}}
\def\hrho{\hat{\rho}}
\def\dchi{\hat{\th}}

\newcommand{\Xm}{{X_{m}}}
\newcommand{\Xmb}{{X_{\overline{m}}}}
\newcommand{\xm}{{x_{m}}}
\newcommand{\xmb}{{x_{\overline{m}}}}

\newcommand{\rrm}{{r_{m}}}
\newcommand{\rmb}{{r_{\overline{m}}}}

\newcommand{\ym}{{y_{m}}}
\newcommand{\ymb}{{y_{\overline{m}}}}

\newcommand\smmb{{\sum_{i=m,\overline{m}}}}

\begin{document}

\title {
Friedel oscillations and horizon charge in 1D holographic liquids}

\preprint{NSF-KITP-12-122}

\author{Thomas Faulkner}
\affiliation{Kavli Institute for Theoretical Physics, University of California, Santa Barbara CA 93106 }

\author{Nabil Iqbal}
\affiliation{Kavli Institute for Theoretical Physics, University of California, Santa Barbara CA 93106 }

\begin{abstract}

In many-body fermionic systems at finite density correlation functions of the density operator exhibit Friedel oscillations at a wavevector that is twice the Fermi momentum. We demonstrate the existence of such Friedel oscillations in a 3d gravity dual to a compressible finite-density state in a (1+1) dimensional field theory. The bulk dynamics is provided by a Maxwell $U(1)$ gauge theory and all the charge is behind a bulk horizon. The bulk gauge theory is compact and so there exist magnetic monopole tunneling events. We compute the effect of these monopoles on holographic density-density correlation functions and demonstrate that they cause Friedel oscillations at a wavevector that directly counts the charge behind the bulk horizon. If the magnetic monopoles are taken to saturate the bulk Dirac quantization condition then the observed Fermi momentum exactly agrees with that predicted by Luttinger's theorem, suggesting some Fermi surface structure associated with the charged horizon. The mechanism is generic and will apply to any charged horizon in three dimensions. Along the way we clarify some aspects of the holographic interpretation of Maxwell electromagnetism in three bulk dimensions and show that perturbations about the charged BTZ black hole exhibit a hydrodynamic sound mode at low temperature. 


\end{abstract}

\today

\maketitle

\tableofcontents


\section{Introduction}

This paper will be concerned with a holographic description of a particular compressible phase of matter. Recall that a compressible state is one in which the density of a $U(1)$ charge is a continuous function of various parameters such as the chemical potential $\mu$ or the temperature $T$. In traditional perturbative quantum field theory, it is generally understood that for such a state to persist to arbitrarily low temperatures without breaking the $U(1)$ symmetry, the charge cannot be carried by bosons, which have a tendency to condense and form a superfluid phase. The charge then must then be carried by fermionic excitations, which will fill out one or more Fermi surfaces. One would then naively expect that in {\it any} compressible state at sufficiently low temperature various correlation functions -- such as those of the charge density itself -- should exhibit sharp singularities in momentum space associated with these Fermi surfaces. 

In fact the locations of these singularities are tightly constrained by Luttinger's theorem \cite{Luttinger}, which relates the volume enclosed by the Fermi surface to the charge density $\rho$ measured in units of the fundamental charge quantum $q_e$. For example, in a $1+1$ dimensional field theory, this relation reads
\be
\rho = \frac{q_e}{2\pi}(2 k_F), \label{normalkf}
\ee
with $k = \pm k_F$ the locations of the Fermi points. In the standard field theoretical examples the density-density correlator $\langle \rho(k) \rho(-k) \rangle$ exhibits singularities at $k = 2k_F$. Upon Fourier transformation the resulting $\cos(2k_F x)$ structures in position space are called Friedel oscillations. We see that these oscillations are actually rather important: they provide a direct probe of the underlying Fermi surface. We note at this point that in (1+1) dimensions the above discussion degenerates slightly in that there is really no fundamental difference between a density of bosons and of fermions: there is no true superfluid phase in one spatial dimension and a finite density of {\it bosons} also exhibits Friedel oscillations at the same wavevector \eqref{normalkf}. Thus the correlation between Friedel oscillations and charge density is even tighter in (1+1) dimensions.  

\subsection{Problem: the nature of horizon charge}

It is thus somewhat perplexing that gauge/gravity duality \cite{AdS/CFT} presents us with many examples of compressible states which at first glance do {\it not} exhibit any such singularities. Here we study a strongly coupled field theory with a large number (``$N^2$'') of degrees of freedom. The dual gravitational description of any such state involves a charged black hole horizon that sources electric flux in the bulk. This flux then penetrates through the bulk to the conformal boundary of the spacetime, where its boundary value can be  interpreted as the charge density $\rho$ of the dual field theory via the usual AdS/CFT dictionary. In some examples this horizon structure persists to zero temperature. Despite extensive study of these systems in applications of holography to condensed matter physics (see e.g. \cite{Hartnoll:2009sz,McGreevy:2009xe,SachdevNew} for reviews), it is generally difficult to identify exactly what field-theoretical degrees of freedom are carrying the charge. The current understanding of such a system is that the charge is carried by gauge-charged excitations in the dual field theory \cite{Huijse:2011hp,Iqbal:2011in,Hartnoll:2011fn,Sachdev:2010um,Sachdev:2010uz}: in the holographic literature such phases are often called ``fractionalized''\footnote{The name arises from an analogy with related systems in condensed matter theory, where any gauge group is emergent and associated with the dissociation of the electron into degrees of freedom with fractional quantum numbers. These fractional excitations should be identified with gauge-charged degrees of freedom in holography.}. Note that in such a system we cannot easily access any correlators corresponding to {\it fundamental} fermions, which presumably are not gauge-invariant. However we {\it can} compute the density-density correlator, and despite extensive study so far no correlators computed in these backgrounds demonstrate any special structure in momentum space \cite{Edalati:2010pn,Edalati:2010hk,Hartnoll:2012wm}\footnote{Note that singularities in momentum space in a holographic model have been found in the recent work \cite{Polchinski:2012nh}; the setup is rather different, involving a density of strings rather than particle-like excitations, and the connection to the results of this paper is not clear.}. Clearly there is some tension with the field-theoretical intuition described above.  

It is the goal of this paper to address this puzzle in the context of a specific model, of the 3d gravitational dual of a 1+1-dimensional quantum liquid. We will demonstrate that if we take into account appropriate {\it non-perturbative} effects in the bulk, the expected singularities are indeed present in the density-density correlation at the expected location in momentum space, although their amplitude is strongly suppressed. This suggests -- as one would have hoped -- that some dual Fermi surface structure exists in these charged black holes, despite the fact that all of our calculations are of gauge field dynamics on a 3d curved background, with no explicit charged fermions in sight. 

We stress that this problem is different from the inclusion of explicit fermions in the {\it bulk}. Such bulk fermions are dual to gauge-invariant charged operators in the field theory. These systems have been extensively studied and often possess explicit Fermi surfaces (\cite{Lee:2008xf,Liu:2009dm,Faulkner:2009wj,Cubrovic:2009ye,Hartnoll:2011dm}; see \cite{Iqbal:2011ae} for a review) in the correlation functions of the dual fermionic operators. These gauge-neutral fermions contribute a parametrically small ($O(1)$ vs. $O(N^2)$) charge density that is outside the black hole horizon. This is dual to the fact that the charge density is carried by gauge-invariant excitations in the field theory and so can essentially be understood in the framework of traditional field theory, as has been emphasized recently by various authors \cite{Sachdev:2011ze,Hartnoll:2011fn,Iqbal:2011bf}. In particular, Friedel oscillations sourced by these explicit bulk fermions have been studied in \cite{Puletti:2011pr}. There are no such bulk fermions in our description.

\subsection{Monopoles and Berry phases}
We now briefly describe the relevant structures in the bulk, leaving a detailed description to the next section. We will study a strongly coupled 1+1 dimensional field theory with a conserved current $j^{a}$. This field theory has a three-dimensional gravitational dual which we take to be weakly coupled. We will assume that the  current $j^a$ is dual to a bulk gauge field $A_{\mu}$, and that the leading dynamics of this gauge field are given by a bulk Maxwell term $(dA)^2$. We note that while this is standard in higher dimensional examples of gauge/gravity duality, generally in AdS$_3$/CFT$_2$ one expects a different structure involving bulk Chern-Simons terms. This is not quite the model we will study, and we will elaborate on the relation between this model and ours in Section \ref{sec:prelim}. 

Turning on a chemical potential for the boundary theory current, we find that the relevant bulk gravity solution is the charged BTZ black hole \cite{btz,Martinez:1999qi}. The detailed structure of this solution is not important. The key fact is that there is a nonzero bulk electric field present, sourced by the horizon and pointing outwards in the emergent holographic direction:
\be
F^{rt}(r) \propto \rho \ . \label{bgEfield}
\ee
The standard AdS/CFT prescription relates fluctuations of the bulk gauge field $A_{\mu}$ about this background to the correlation functions of the boundary theory current $j^a$. As mentioned above, correlation functions previously calculated using this method do not reveal any nontrivial structure in momentum space. This is perhaps not surprising. If we take the relation \eqref{normalkf} seriously, we see that the location of any such singularity -- i.e. the {\it value} of  $2k_F$ -- depends not only on $\rho$ but also on $q_e$, the charge of a single quantum excitation in the field theory. The black hole horizon and linearized fluctuations around it know nothing of any $q_e$, and so it seems clear that they cannot reproduce this answer. 

However, $q_e$ does have a bulk interpretation. It is expected that in any theory of quantum gravity -- such as the one that we are studying in the bulk -- all gauge symmetries should be {\it compact}. In particular $A_M$ should be a compact $U(1)$ gauge field with a minimum quantum of charge, and it is easy to see that this charge quantum should be identified with the quantum of charge characterizing the boundary theory Hilbert space. Thus the definition of the bulk theory does contain a $q_e$, even if the black hole solution does not seem to make explicit use of it. 

\begin{figure}[h]
\begin{center}
\includegraphics[scale=0.6]{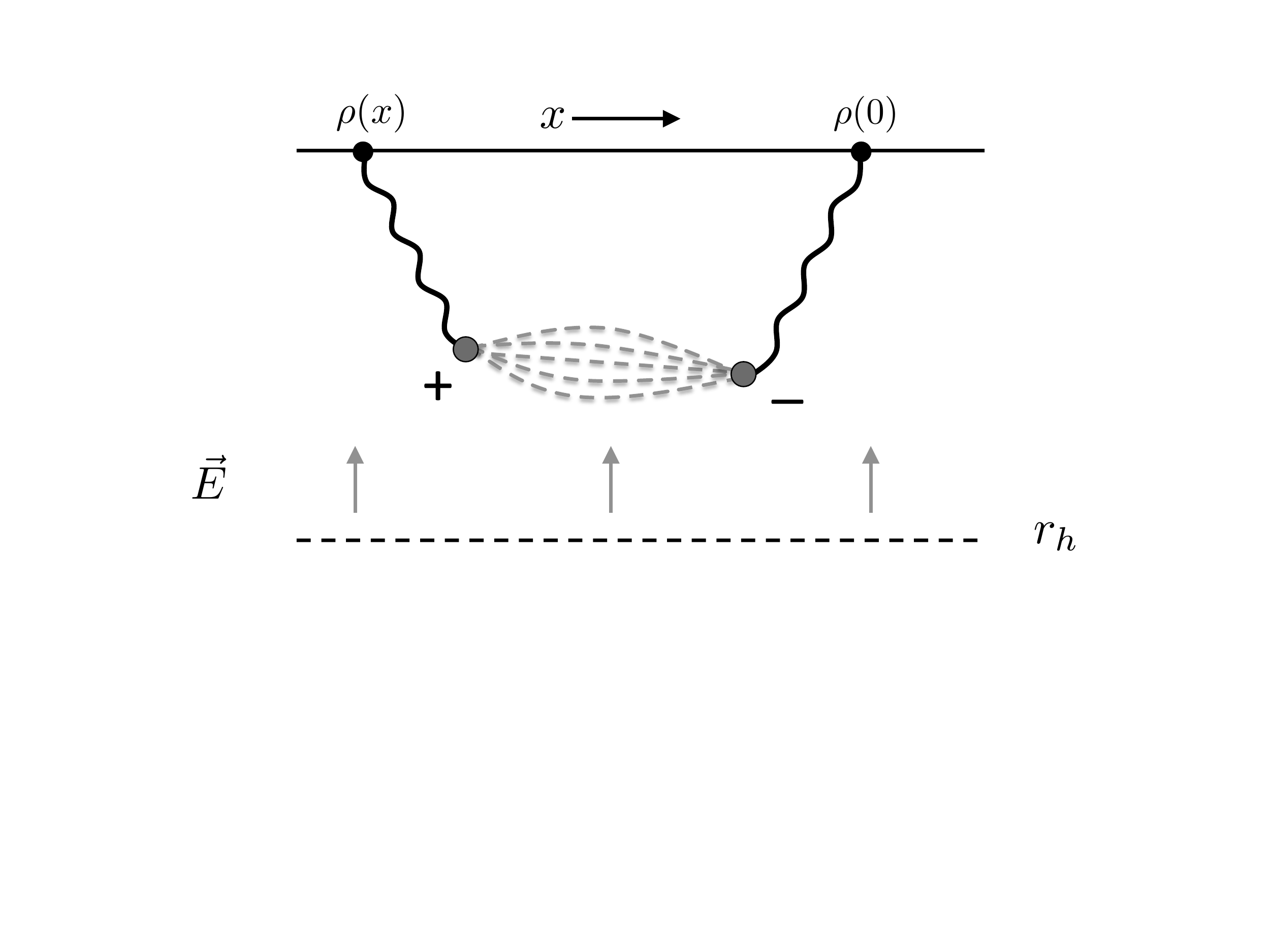}
\end{center}
\vskip -0.5cm
\caption{Witten diagram used in calculation of contribution of monopole-anti-monopole pair to boundary theory density-density correlator. Solid wiggly lines indicate bulk-to-boundary correlators, and dotted gray lines indicate monopole-antimonopole interaction energy.}
\label{fig:witten1}
\end{figure}

Once we accept that the bulk gauge theory is compact, however, a new ingredient presents itself: {\it magnetic monopoles}. In three Euclidean dimensions these objects are localized in both space and time, and so should be thought of as instantons. If a monopole with magnetic charge $q_m$ is placed at a point $x_m$ then we find that magnetic flux is created there, i.e.
\be
dF(X) = q_m \delta^{(3)}(X-X_m)
\ee
On general grounds we expect these monopoles to represent allowed tunneling events in the theory. They can have a profound effect on infrared physics; for example in flat space in three dimensions they drive confinement of compact $U(1)$ gauge theory \cite{Polyakov:1976fu}. In the remainder of this paper we will compute the effect of bulk monopoles through AdS/CFT on the density-density correlation function of the dual field theory. The calculation is conceptually quite simple: monopoles in the bulk source the bulk gauge fields, resulting in a nontrivial contribution to the boundary theory density-density correlators. As we explicitly demonstrate, the problem can be formulated in terms of a Witten diagram as in Figure \ref{fig:witten1}, where the monopoles are integrated all throughout the bulk with the appropriate action cost. We will see that these monopoles provide a nontrivial probe of the charge density.

In particular, the boundary theory charge density manifests itself as the bulk background electric field \eqref{bgEfield}. In the presence of such an electric field, each monopole event is weighted by a nontrivial {\it phase}, which we will call a Berry phase in a slight abuse of notation. As we will explain in detail later, this phase takes the form
\be
S_B(X_m) = i q_m \rho \x_m 
\ee
where $x_m$ is the spatial coordinate of the monopole in the direction parallel to the horizon. This is completely analogous to the Aharonov-Bohm phase accumulated by an electric charge in a background magnetic field. We see that the phase associated with the monopole oscillates in space with period $\rho q_m$, indicating a special feature at a particular position in momentum space. This feature turns out to be the expected Friedel oscillation. The existence of this phase is quite easy to understand: the bulk of the paper is devoted to determining the form of the interaction energy between monopoles, which is crucial in determining the precise form of the singularity. Similar Berry phases associated with monopole events play an important role in traditional condensed matter physics (see e.g. \cite{sachdevread1, sachdevread2}). 

Comparing to the field-theoretical expression \eqref{normalkf} we see that there is a nontrivial interplay between electric and magnetic charges. Fascinatingly, we will find precise agreement between the field-theoretical and gravity expressions if the spectrum of magnetic and electric charges in the bulk {\it saturates} the Dirac quantization condition, i.e. $q_e q_m = 2\pi$. We comment on this (and associated issues) in the discussion section.  

We now briefly summarize the paper. In Section \ref{sec:prelim} we define the model that we will be studying -- Maxwell electromagnetism on AdS$_3$ and discuss some tree-level aspects: in particular we explain how it can be understood as a two-dimensional field theory with a conserved current and a logarithmically marginal deformation. In Section \ref{sec:monsum} we develop a formalism for computing monopole instanton corrections to holographic correlators. In Section \ref{sec:highT} we compute this correction and display the Friedel oscillations in the high-temperature limit, in which gravity decouples from the gauge field, greatly simplifying the calculation. 

In Section \ref{sec:zeroT} we turn to the low temperature limit, in which the mixing with gravity is crucial. Here we first study the system at tree level and exhibit a sound mode in fluctuations about the charged BTZ black hole. In addition to being interesting in its own right, this sound mode determines the long-range correlations between monopoles and so is needed for the instanton calculation. We then present two routes to the zero-temperature Friedel oscillations: Section \ref{friedmatch} is very quick, but we feel that it obscures some important physics involving the interaction of monopoles with gravity, which we develop in detail in Section \ref{sec:3dgrav}. We summarize our results and discuss some implications in Section \ref{sec:conc}. The trusting and impatient reader can skip to the conclusion section with little loss of continuity.

\section{Preliminaries} \label{sec:prelim}
In this section we define our model and establish some facts that will be needed to derive the Friedel oscillations. We will study a 1+1 dimensional field theory with a conserved current $j^{a}$. We assume that this theory has a weakly coupled gravity dual governed by the classical Euclidean action
\be
S_E =  \int d^3 X\sqrt{g}\le(\frac{1}{16 \pi G_N}\le(-R - 2\ri) + \frac{1}{4g_F^2} F^2\ri), \label{3daction}
\ee
where $A_{\mu}$ is the bulk gauge field dual to the boundary theory current and $F = dA$. $\mu,\nu$ run over the three bulk directions, and $a,b$ will run over boundary theory spacetime coordinates. In general we will use $y$ to refer to boundary theory coordinates, so $X^{\mu} = (r, y^a) = (r,\tau,x)$\footnote{Our conventions for the Levi-Civita tensor in 2d and 3d are
\be
\ep^{xr\tau} = \frac{1}{\sqrt{g}} \qquad \ep^{rab} = \frac{1}{\sqrt{g}}\ep^{ab}.
\ee
Two-index epsilon always lives in flat 2d space and refers to the symbol, not the tensor.}.
 We will work with spacetimes which are asymptotically AdS, taking the form
\be
ds^2 = r^2\le(d\tau^2 + dx^2\ri) + \frac{dr^2}{r^2}\,  \label{eads3}
\ee 
as $r \to \infty$. The AdS radius has been set to $1$ for notational convenience. We will generally work in Euclidean signature but will sometimes Wick rotate to real time via
\be
t = -i\tau \qquad \om = i \om_E \ \qquad iS = -S_E \ .
\ee 

In gauge/gravity duality in higher dimension an action such as \eqref{3daction} is standard for the bulk dual of a field theory with a conserved current. However in two dimensions the story is usually different. A conserved current in a two-dimensional conformal theory is associated with a higher symmetry: Ward identities guarantee the separate conservation of both the holomorphic and antiholomorphic part of the current $j^{a}$, resulting in an extended current algebra. This current algebra is dual to a doubled Chern-Simons theory in the 3d bulk, with one Chern-Simons gauge field for the holomorphic and anti-holomorphic part each (see e.g. \cite{Kraus:2006wn}). Thus the action \eqref{3daction} may appear somewhat unfamiliar, and though it has been studied in a number of recent works its correct holographic interpretation has been debated \cite{Jensen:2010em}. 

We claim that the action \eqref{3daction} can actually be holographically interpreted as the dual to a two-dimensional field theory with a dynamical conserved current. However the theory is not conformal: it has a marginal deformation, which is simply $j^{a}j_{a}$ and which must be dealt with appropriately. We will discuss the appropriate boundary conditions below: however first we will use scalar/vector duality in 3d to reformulate the bulk dynamics in terms of a scalar $\Th$ rather than the gauge field $A_{\mu}$. This is very convenient for dealing with monopoles, which are simply point sources for the field $\Th$. 

\subsection{Dualizing gauge fields in the bulk and AdS/CFT} \label{sec:doppel}
We begin with the bulk gauge field action in Euclidean signature
\be
S_A = \int d^{3}X \sqrt{g} \frac{1}{4g_F^2} F^2 \ . \label{gaugeac}
\ee
It is well-known that in three bulk dimensions the dynamics of a $U(1)$ gauge field is identical to that of a scalar via $dA \sim \star d\Th$. To see this at the level of the path integral we note that it is implicit in \eqref{gaugeac} that the dynamical variable is the gauge field $A$, i.e.
\be
Z_{bulk} = \int [\sD A] \exp\le(-S_A\ri) \ .
\ee
Consider replacing the integral over $A$ with one over the field strength $F$ instead. In that case the Bianchi identity $dF = 0$ will no longer be explicitly satisfied, and so we should introduce a Lagrange multiplier field $\Th$ to enforce it:
\be
Z_{bulk} = \int [\sD F] [\sD \Th] \exp\le(-S_A + \frac{i}{2}\int d^{3}X \sqrt{g}\;\Th(X) \ep^{\mu\nu\rho}\p_{\mu}F_{\nu\rho}(X)\ri) \label{mixed}
\ee
The normalization of $\Th$ is arbitrary at this point: we shall explain the rationale for the factor of $i$ later. The action for $F$ is quadratic, and we can integrate it out by imposing its equation of motion in the action,
\be
\ep^{\mu\nu\rho} \p_{\mu} \Th = \frac{i}{g_F^2}F^{\nu\rho}, \label{chiFrel}
\ee
after which we find simply
\be
Z_{bulk} = \int [\sD \Th] \exp\le( -\frac{g_F^2}{2}\int d^{3}X \sqrt{g}(\nabla \Th)^2\ri), \ee
i.e. the action of a free massless scalar. Note that the dynamical Maxwell equation $d \star F = 0$ has become an identity in terms of $\Th$, while the Bianchi identity $dF = 0$ is equivalent to the dynamical scalar equation of motion $d \star d\Th = 0$. 

The preceding manipulations are entirely standard. Now we should interpret them from the point of view of AdS/CFT. Usually we interpret the boundary value of the gauge field $A_{\mu}(r\to\infty)$ as determining the source for the boundary theory current $j^a$, i.e
\be
Z_{bulk}\big|_{A(r\to\infty) = a} = \left\langle \exp\le(\int d^2y\;a_a j^a\ri)\ri\rangle_{QFT} \ . \label{bulkbdy}
\ee
To write this boundary condition in terms of the bulk scalar we note that a conserved current in two dimensions can be related to a scalar operator $\sO$ as

\be
\label{opo}
j^a = i \ep^{ab}\p_b \sO \ .
\ee
where the $i$ is natural if we are working in Euclidean signature.  We insert this on the right-hand side of \eqref{bulkbdy} and integrate by parts, after which the source term becomes
\be
\int d^2x \le(i \ep^{ab}\p_a a_b \ri) \sO . 
\ee
Thus the operator $\sO$ naturally couples to the field strength of $a$, which in two dimensions is a scalar. However in AdS/CFT the field strength of $a$ is precisely the boundary value of the bulk field strength $F_{\mu\nu}(r \to \infty)$, which from \eqref{chiFrel} is:
\be
\frac{i}{2} \ep^{ab}F_{ab} = g_F^2 \sqrt{g}g^{rr}\p_r \Th \label{Fpirel}
\ee
The right hand side is 
the bulk canonical momentum $\Pi$ conjugate to $\Th$ with respect to an $r$-foliation. Thus in the scalar representation it is the boundary value of $\Pi$ that is the source for the field theory operator $\sO$. Note that a massless scalar $\Th$ in AdS$_3$ has two possible quantizations with different dimensions of the boundary theory operator $\Delta_{\pm} = 1 \pm 1$ \cite{Klebanov:1999tb}; choosing $\Pi$ as the source picks the dimension $\Delta_- = 0$, which is indeed the dimension of $\sO$. Note that dualizing a gauge field in $3+1$ bulk dimensions has a similar
action on the boundary conditions of the dual gauge field \cite{Witten:2003ya, Marolf:2006nd}.

Thus to compute two-point functions of the current $j^a$ from AdS/CFT in the scalar representation, we should use the standard holographic prescription to compute boundary correlation functions of the operator $\sO$ dual to $\Th$ with the quantization appropriate to $\Delta_-$ and then use
\be
\langle j^a(y_1) j^b(y_2) \rangle =  -\ep^{ac} \ep^{bd} \frac{\p}{\p y_1^c} \frac{\p}{\p y_2^d} \langle \sO(y_1) \sO(y_2) \rangle \label{currcurr}. 
\ee
There is thus only one nontrivial function's worth of information in the current-current correlator: in two dimensions current conservation strongly constrains the correlator. Note also that using the standard AdS/CFT expression for the boundary theory current
\be
\langle j^a \rangle = \frac{1}{g_F^2}\sqrt{g}F^{ar}(r \to \infty),
\ee 
we can express the current in terms of the scalar using \eqref{chiFrel}:
\be
\langle j^a \rangle= i\ep^{ab}\p_b \Th\bigg|_{r \to \infty} \ . \label{currscalar}
\ee
Thus local conservation of the current is now an identity. Comparing this to \eqref{opo} wee
see that $\langle \sO \rangle = \Th |_{r\to \infty}$ which makes it clear we are allowing $\Th$ to fluctuate on the boundary. 

In Lorentzian signature we  have $\langle j^t \rangle = \p_{x}\Th(r \to \infty)$. If we consider defining the field theory on a circle by identifying $x \sim x + L_x$, then the total field theory charge $Q = \int_0^{L_x} dx \langle j^t \rangle$ is simply the total winding in $\Th$ at infinity. In particular, the requirement that the field theory charge be quantized in units of $q_e$ means that $\Th$ should be a periodic variable, i.e. $\Th \sim \Th + q_e$.  
\subsection{Marginal deformations and boundary conditions}\label{sec:bc}
We have now understood a scalar representation of pure Maxwell theory on AdS$_3$. We previously mentioned this theory can be understood as being holographically dual to a two-dimensional field theory with a conserved current and a marginal deformation $j^a j_a$. This deformation is not exactly marginal; depending on its sign it is either marginally relevant or irrelevant, resulting in logarithms in various expressions. We now demonstrate these facts with a systematic study of the appropriate boundary conditions on bulk fields. For convenience we use the scalar representation derived above. 


For this section we will work on pure Euclidean AdS$_3$ with metric \eqref{eads3}.
It will be helpful to work with the canonical momentum $\Pi$ conjugate to $\Th$, which we take to be
\be
\Pi = -\sqrt{g}g^{rr}\p_r \Th \ . 
\ee
We work in Fourier space with
\be
\label{fconv}
\Th(r,x,\tau) = \Th(r)e^{i k x + i \om \tau} \qquad p^2 \equiv \om^2 + k^2,
\ee
and the massless scalar equation of motion is then
\be
\p_r\le(r^3 \p_r \Th\ri) - \frac{p^2}{r} \Th = 0.
\ee
A near-boundary analysis of the solutions to this equation yields:
\be
\Th(r \to \infty) \sim A + \frac{1}{r^2}\le(B - A \frac{p^2}{2} \log r\ri) .\label{chiexp}
\ee

There are two possible boundary conditions on $\Th$. In one of them, we take the boundary value of $\Th(r \to \infty)$ itself to be the source, i.e. standard quantization. This is a conformally invariant boundary condition, as the logarithmic term in \eqref{chiexp} is subleading at large $r$. It corresponds to the dual operator $\sO$ having a conformal dimension $\Delta_+ = 2$. Following the discussion of the previous section this quantization does not correspond to a dynamical current. Indeed from \eqref{currscalar} we see that fixing the boundary value of the scalar corresponds to fixing the field-theory current, and thus this quantization corresponds to having a {\it dynamical} $U(1)$ gauge field on the boundary, as is discussed in \cite{Marolf:2006nd, Jensen:2010em}.

To obtain a dynamical {\it current} we thus have to choose the other quantization, corresponding loosely to fixing the value of the canonical momentum $\Pi$ at infinity, which via \eqref{Fpirel} is equivalent to specifying the boundary electric field $F_{x\tau}$ that the fluctuating current feels. This corresponds to the ``conformal'' dimension $\Delta_- = 0$. However the logarithmic term 
in \eqref{chiexp} means that $\Pi$ at infinity actually changes slowly with $r$, never approaching a constant. Thus there is an ambiguity in what we mean by ``the boundary value of $\Pi$''. From \eqref{chiexp} at infinity we have
\be
\frac{d}{d \log r}\Pi(r) = -p^2 \Th(r)
\ee
Thus any boundary condition cannot be imposed at infinity and must be imposed at some specific value of $r$ that defines a cutoff, say $r = r_{\Lam}$. Consider then the following general boundary condition for a normalizable perturbation:
\be
\ka \Pi(r_{\Lam}) = p^2 \Th(r_{\Lam}) \label{bc1} \ . 
\ee 
Note that we are fixing a linear combination of $\Pi$ and $\Th$ at infinity, and $\ka$ is thus a double-trace coupling, which runs logarithmically \cite{Witten:2001ua}. To understand this consider instead imposing the boundary condition at a different value of $r_{\Lam}' \equiv \lam r_{\Lam}$ with a different value for the coupling $\ka'$. Then from the expansion we have $\Pi(r_{\Lam}') = \Pi(r_{\Lam}) - \Th p^2 \log \lam$, and thus the {\it new} boundary condition can be written
\be
\frac{\ka'}{1 + \ka' \log \lam}\Pi(r_{\Lam}) = p^2 \Th(r_{\Lam})\ .
\ee
This is actually {\it equivalent} to the original boundary condition \eqref{bc1} if $\ka$ and $\ka'$ are related by
\be
\frac{1}{\ka'} = \frac{1}{\ka} - \log \lam \label{rg1} \ .
\ee
This is precisely the behavior of a logarithmically running coupling. 

To have a well-defined problem we specify the value of $\ka$ at some scale; to understand the physics at a different scale we can use the evolution law given by \eqref{rg1}. As usual, physical observables should not depend on $\ka$ and $r_{\Lam}$ individually: rather, dimensional transmutation occurs, and all physical observables can be expressed in terms of the RG-invariant scale $r_{\star}$:
\be
r_{\star} = r_{\Lam} e^{\frac{1}{\ka}} \ . \label{rstardef}
\ee
This scale should be viewed as the strong-coupling scale of the problem. Note for $\ka > 0$ we have $r_{\star} > r_{\Lam}$, and the coupling is marginally irrelevant; if we study only infrared observables then its effect is benign, only introducing logarithms in various expressions. This is the case we will study in this paper. 

If $\ka < 0$, then $r_{\star} < r_{\Lam}$ and this coupling is growing strong in the infrared, probably resulting in new infrared physics. While we do expect this to be interesting, we will not study this case further in this work. Note that $\ka = 0$ separates two theories with very different infrared physics and so formally corresponds to a quantum critical point, more specifically one of the type called ``marginal'' in \cite{Iqbal:2011aj}.

We now interpret the boundary condition \eqref{bc1} in terms of the conserved current. Using \eqref{currscalar} and \eqref{Fpirel} we have
\be
\ep^{ab} \p_{a}\langle j_b \rangle = \frac{1}{2 g_F^2} \ka \ep^{ab}F_{ab}(r_{\Lam}).
\ee
This expression may appear familiar: the left-hand side is the divergence of the {\it axial} current $\ep^{ab}j_b$ in the field theory, and $F_{ab}$ is the field-theory source. In a 2d {\it conformal} field theory an expression such as this is true at {\it all} scales, with $\frac{\ka}{g_F^2}$ replaced by a precise integer, the level of the current algebra $k$. Indeed this expression would have been the axial anomaly in 2d, and should be thought of as a Ward identity that allows a current $j^a$ to be split into separately conserved holomorphic and antiholomorphic parts. In our theory, however, this expression is only true at a particular $r$-value. This is dual to the fact that $\ka$ requires renormalization, spoiling the current algebra structure. This logarithmic running of $\ka$ is the crucial difference between Maxwell electromagnetism in the bulk (which is dual to only a single conserved current) and a doubled Chern-Simons theory (which is dual to a Kac-Moody current algebra). We will comment further on the relation between these two theories in the conclusion; for now we will simply study the Maxwell theory. 

Note that a boundary action which is consistent the boundary condition \eqref{bc1}
can be constructed and 
is given in Appendix~\ref{sec:bdry}. It contains a term $S_{\ka}$ which we can directly interpret as the current
double trace coupling:
\be
\label{dbltrace}
S_{\ka} = - \frac{g_F^2}{2\ka} \int_{r = r_\Lambda} d^2 y (\nabla \Th)^2 = 
\frac{g_F^2}{2\ka} \int_{r= r_\Lambda} j^a j^a \ ,
\ee
where we have used \eqref{currscalar}. 

For future reference we note that if want to apply a field theory source $H(y)$ 
for the operator $\mathcal{O}$ defined in \eqref{opo} then the appropriate boundary condition is (from \eqref{bc1}):
\be
\Pi(r_{\Lam},y) + \frac{1}{\ka}\nabla^2 \Th(r_{\Lam},y) = H(y), \label{bcsource}
\ee
where $y$ only runs over field theory directions. 

Using \eqref{currcurr} we conclude that the current-current correlation function in momentum space is:
\be
\label{jjcorr}
\langle j^a(p) j^b(-p) \rangle  = \frac{1}{g_F^2} \left. \frac{\epsilon^{ac} \epsilon^{bd} p_c p_d}{ \left(\Pi/\Th\right) - \frac{1}{\kappa}p^2 } \right|_{r=r_\Lambda}
\ee
For illustrative purposes we compute this two-point function for pure AdS$_3$ \eqref{eads3}. The solution that is regular in the interior is    a Bessel function $\Th(r) = r^{-1}K_1\le(\frac{p}{r}\ri)$; expanding this at infinity we find
\be
\langle j^a(p) j^b(-p) \rangle = \frac{1}{g_F^2} \frac{\epsilon^{ac} \epsilon^{bd} p_c p_d}{p^2 \log\le(\frac{p}{\bar{r}_{\star}}\ri)} \qquad \bar{r}_{\star} = 2 r_{\star} e^{-\ga}, \label{currads3}
\ee   
As claimed $r_{\Lam}$ and $\ka$ have combined (modulo factors of 2 and the Euler-Gamma number) into the RG-invariant combination $r_{\star}$. The correlator runs logarithmically with momentum: a CFT correlator takes exactly the same form except with the logarithm replaced by a constant. 

Note that the correlator appears to have a pole at $p^2 = \bar{r}_{\star}^2$. 
From \eqref{rstardef} for an irrelevant coupling $\ka > 0$ this pole lies at $r_{\star} \gg r_{\Lam}$: physically this pole is at a momentum far beyond the cutoff and is spurious. In fact even the mathematical analysis leading to it is incorrect, as the expansion of the Bessel function used to obtain \eqref{currads3} is not valid for $p \gg r_{\Lam}$. Such singularities, which
turn out to be ghosts, were discussed carefully in \cite{Andrade:2011dg} and their removal by an appropriate UV completion is discussed in \cite{Andrade:2011aa}.
\subsection{Charged BTZ black hole} \label{chbtzsec}

We will now write down a black hole solution of (\ref{3daction}) with an electric flux
emanating from the horizon. This is the charged BTZ black hole \cite{btz,Martinez:1999qi}. Similar states have been studied in the context of applications of holography to condensed matter in \cite{Ren:2010ha, Hung:2009qk, Gao:2012yw}. 
It defines a field theory state at nonzero charge density, which will be the subject of our
investigation. The metric and gauge field are:
\be
\label{chbtz}
ds^2 = f(r) d\tau^2 + r^2 dx^2 + \frac{dr^2}{f(r)} \,, \qquad A = i \left(\frac{ g_F^2}{16 \pi G_N}\right)^{1/2} 
\hat{\rho} \log (r/r_h) d\tau \, , 
\ee
where  $f(r) = r^2 - r_h^2 -  (\hat{\rho}^2/2) \log(r/r_h)$. The factor of $i$ in $A$ is there because
we are working in Euclidean. 
It is useful to think
of $\hat{\rho}$ as the energy scale in the field theory below which the charge density
becomes dynamically important. The real charge density is related to this parameter
by a large factor (of order ``$N^2$''):
\be
\label{jt}
\langle j^t \rangle= \frac{i}{g_F^2} \sqrt{g}F^{r \tau}   \equiv \rho  =  
\hat{\rho} \left( \frac{1}{16 \pi G_N g_F^2} \right)^{1/2} \, .
\ee
We can also write the solution in terms of the dual scalar where non-zero
charge density imposes a boundary condition on the winding of
$\Th(\x+L_\x) = \Th(\x) + \rho L_\x$ around the $\x$ direction. Using \eqref{chiFrel} the required
solution is then simply:
\be
\Th = \rho \x
\ee
This fact -- that a radial electric field corresponds to a dual scalar that depends on the direction parallel to the horizon -- is the ultimate origin of the Friedel oscillation structure in the boundary field theory correlators. 
 
The black hole thermodynamics can be worked out. The horizon lies
at $r=r_h$ and the Hawking
temperature and chemical potential are:
\be
\label{thermo}
T = \frac{ r_h}{2\pi}\left (1 - \frac{\hat{\rho}^2}{4r_h^2}\right)\, , \qquad \mu =  \left(\frac{ g_F^2}{16 \pi G_N}\right)^{1/2}   \hat{\rho} \log(r_\star/ r_h) \, .
\ee
In computing the chemical potential we have included a contribution, on top of the usual
AdS/CFT answer, from the double trace current deformation given in \eqref{dbltrace}:
\be
\mu = A_t(r_\Lambda) +  \frac{  g_F^2}{2 \kappa} \frac{\delta}{\delta j^t} \langle j^t j^t \rangle
 = A_t(r_\Lambda) + \frac{g_F^2}{\kappa} \rho
\ee
As expected the two contributions combine to give the RG invariant $\log(r_\star/r_h)$
in \eqref{thermo}.
The zero temperature limit reveals an extremal horizon at $r_h = 2 \hat{\rho}$ with
non-zero entropy density $S \propto \hat{\rho}/G_N$
and an associated AdS$_2 \times \mathbb{R}$ near horizon geometry. These specific
features of the zero-$T$ limit
will \emph{not} be important to our results which will continue to
hold in other more general backgrounds with charged horizons: for example
one could consider three dimensional versions of the backgrounds discussed
in \cite{Goldstein:2009cv,Charmousis:2010zz,
Hartnoll:2011pp}.

More importantly, there are various instabilities associated with 
the horizon radius becoming too large compared to the UV scale $r_\star$.
For example at $T=0$ the compressibility is
\be
\label{susc}
\frac{d\rho}{d\mu} = \frac{1}{g_F^2}\frac{1}{\log(r_\star/r_h) - 1},
\ee
which diverges at $r_h = r_\star/e$ and is negative for larger $r_h$.
A negative susceptibility indicates a thermodynamic instability. 
This is not surprising: we should UV complete this theory at
some energy/radius below the scale $r_\star$ and we do not expect black holes to make
any sense in our current setup beyond this radius. 
We note that this is also an obstruction
to making sense of the theory when $r_\star$ is an IR scale
perhaps associated with some strong dynamics. That is when $\kappa < 0$ such that
$r_\star < r_\Lambda$ from \eqref{rstardef}.
As discussed above, in this paper we will only study the case
when $r_{\star}$ is a UV scale, and so we will always have $r_{\star} \gg r_h$
with stable thermodynamics.

Finally, if we consider a probe limit by taking $G_N \to 0$ while holding $g_F^2$ fixed, then $\hat{\rho} \to 0$. As the theory is conformal (modulo logarithms) the dimensionless parameter characterizing the temperature is $\frac{T}{\hat{\rho}}$, which goes to infinity in this limit. Thus, as is usual in AdS/CFT, the probe limit corresponds to a high temperature limit. We will work in this probe limit in Section \ref{sec:highT}. 
\section{Monopole sums in AdS} \label{sec:monsum}
We now describe the nonperturbative objects of interest in this paper: magnetic monopoles. We first review some basic features, and then we build the framework for summing over monopoles in AdS and extracting corrections to boundary theory correlation functions. 

\subsection{Monopole generalities}
In three Euclidean spacetime dimensions a monopole is a localized instanton that creates magnetic flux. Near the location of the monopole at $X_m$ we have
\be
\ep^{\mu\nu\rho}\p_{\mu}F_{\nu\rho} = \frac{2 q_m}{\sqrt{g}} \delta^{(3)}(X-X_m), \label{monopoledef}
\ee
where $q_m$ is the magnetic charge of a single monopole. Monopoles exist only in compact $U(1)$ gauge theories, i.e. they must satisfy the Dirac quantization condition
\be
q_e q_m = 2 n \pi, \qquad n \in \mathbb{Z}
\ee
with $q_e$ the unit of electric charge in the theory. These instantons correspond to tunneling events in which the total amount of magnetic flux in the system
\be
\Phi = \int_\Sigma F
\ee
(with $\Sigma$ a spacelike slice) is changed by $q_m$. It is clear from \eqref{monopoledef} that the description of the field strength in terms of a gauge field $F = dA$ is breaking down at the location of the monopole; however the scalar representation introduced in Section \ref{sec:doppel} is ideal for treating monopoles, as we now review. 

We begin by considering the partition function $Z_{bulk}$ in the presence of a single monopole at $X_m$. Recall the scalar $\Th$ was introduced as a Lagrange multiplier to enforce the Bianchi identity. However we no longer want $dF = 0$ everywhere; rather we need to incorporate the monopole source as in \eqref{monopoledef}.  \eqref{mixed} is then modified to read
\begin{gather}
Z_{bulk}[X_m] = \int [\sD F] [\sD \Th] \exp\bigg(-S_A -S_{c} + \frac{i}{2}\int d^{3}X \sqrt{g}\;\Th(X) \big(\ep^{\mu\nu\rho}\p_{\mu}F_{\nu\rho}(X) \\ - \frac{2 q_m}{\sqrt{g}}\delta^{(3)}(X-X_m)\big)\bigg) \label{mixed2},
\end{gather}
where $S_{c}$ is understood to be the action cost associated with the core of the monopole. Eliminating $F$ we find
\be
Z_{bulk}[X_m] = \int [\sD \Th] \exp\le( -\frac{g_F^2}{2}\int d^{3}X \sqrt{g}(\nabla \Th)^2 - S_{c} - iq_m \Th(X_m) \ri),  \label{mbulkac}
\ee
The last term is the coupling of the field $\Th$ to a heavy magnetic source located at a specified point $X_m$. Note this is the magnetic equivalent of the coupling of the gauge field to a heavy electric source $i q_e \int_C A$ located along a specified worldline $C$. The dynamical equation of motion following from varying this action
\be
g_F^2 \nabla^2\Th(X) = iq_m \frac{\delta^{(3)}(X-X_m)}{\sqrt{g}} \label{moneq}
\ee
is equivalent to \eqref{monopoledef}. This action will be our starting point for subsequent calculations. 

It is convenient at this point to introduce the bulk-to-bulk scalar propagator $G$, which satisfies
\be
\nabla^2 G(X,X') = \frac{1}{\sqrt{g}}\delta^{(3)}(X-X') \ . \label{bulkbulkdef}
\ee
The field measured at $X$ produced by a monopole at $X_m$ is $\Th_{X_M}(X) = \frac{i q_m}{g_F^2} G(X,X_M)$. We will construct this propagator explicitly on various geometries in future sections.

The preceding considerations are entirely standard. However we are interested in studying monopoles not in vacuum but on the charged black hole background \eqref{chbtz}. We have a background electric field and so the field $\Th$ has a background value $\Th_0 = \rho x$, i.e. it is linearly increasing in the direction parallel to the horizon. The source term in \eqref{mbulkac} now means each monopole contributes an extra term to the action. Evaluating the source term on the background, we find
\be
S_{berry}[X_m] = -iq_m \Th_0(X_m) = -iq_m \rho \x_m \ . \label{berry}
\ee
This phase acquired by a magnetic charge in an electric field is entirely analogous to the usual Aharonov-Bohm phase accrued by an electric charge moving in a magnetic field, which is found by evaluating $iq_e\int_C A$ along the trajectory of the charge. In the monopole sum, each monopole event should be weighted by such a phase in the functional integral; each monopole truly knows where it is along the $x$ direction. We discuss some elementary aspects related to this phase in Appendix \ref{sec:berry}.

\subsection{On-shell action and Witten diagram}
Now we perform the sum over monopoles. Before starting we note that we assume throughout that the field $\Th$ has a background value $\Th_0 = \rho x$ from the background electric field. The calculations in this section will concern perturbations around this background, i.e. we split
\be
\Th = \Th_0 + \th(X) 
\ee
and deal with the perturbation field $\th(X)$. As $\Th_0$ solves the background equations of motion the action for the perturbation is essentially quadratic, except for a linear contribution from the Berry phase term \eqref{berry} at the location of the monopole sources, which we will explicitly include. 

Thus our starting point is the action for a configuration of a total of $N$ monopoles and $\bar{N}$ anti-monopoles located at points $\{X_i\}$ \eqref{mbulkac}:
\be
S^{(N,\bar{N})}[\{X_i\}] = \int d^3X \sqrt{g}\le(\frac{g_F^2}{2}(\nabla \th)^2\ri) + \sum_{X_i}\le(S_c \pm i q_m \th(X_i) +  S_{berry}[X_i]\ri)
\ee
Here the choice of sign in $\pm$ depends on whether we have a monopole or anti-monopole at the corresponding point. Finally, to enforce a good variational principle the action requires boundary terms that we have not explicitly written down; these will be discussed below. The full bulk partition function can be written as a sum over different monopole sectors, i.e.
\be
Z = Z^{(0,0)} + Z^{(1,0)} + Z^{(0,1)} + Z^{(1,1)} + \dots
\ee
The term $Z^{(0,0)}$ has no monopoles and is the usual perturbative AdS/CFT answer. Here we will compute $Z^{(1,1)}$, corresponding to a single monopole-anti-monopole pair. This is the leading correction; terms with net monopole charge will not contribute to this particular observable. We have
\be
Z^{(1,1)} = \int [\sD \th] \Lambda^6 e^{-2S_c} \int d^3\Xm d^3\Xmb \sqrt{g(\Xm)} \sqrt{g(\Xmb)} \exp[-S[\Xm,\Xmb]] \label{z11}
\ee 
We place the monopole at $\Xm$ and the anti-monopole at $\Xmb$. To determine the full partition function we integrate over their locations, assuming that the correct measure is simply the proper volume with respect to the bulk metric, with the scale provided by $\Lambda$, which we take to be a {\it bulk} UV scale associated with the core of the monopole. Finally, as always in AdS/CFT we need to specify boundary conditions on the fields. We allow for a nontrivial source $H(y)$, and the precise boundary condition at infinity is \eqref{bcsource}. 

We will now evaluate this functional integral by saddle-point. We should thus determine the on-shell action on a monopole configuration. The equation of motion for the field $\th$ is
\be
g_F^2 \nabla^2 \th = \sum_{X_i} \le(\pm\frac{iq_m}{\sqrt{g}}\delta^{(3)}(X-X_i)\ri) \ . \label{chisources}
\ee
Integrating the action by parts and evaluating it on-shell we find for the total classical action:
\be
S_{onshell}[\{X_i\}] =\sum_{x_i} \le(S_c \pm q_m \frac{i}{2}\th(X_i)\ri) + S_{berry}[\{X_i\}] + S_\p[\th(r_{\Lam})], \label{sonsh}
\ee 
where $S_{\p}[\th]$ is a boundary term whose form we will describe below. 

Now we would like to compute the current-current correlator in the boundary theory. Via \eqref{currcurr} this is simply related to the correlator of the operator $\sO$ dual to the field $\Th$, whose two-point function is
\be
g_F^4 \langle \sO(y_1) \sO(y_2) \rangle \equiv \sG(y_1,y_2) = \frac{1}{Z[0]}\frac{\delta^{2}}{\delta H(y_1) \delta H(y_2)} Z[H(y)]
\ee
We are only interested in the contribution from $Z^{(1,1)}$. Putting the on-shell action into \eqref{z11} we find that the correction to the correlator yields
\be
\sG(y_1,y_2) = \frac{\delta^{2}}{\delta H(y_1) \delta H(y_2)} \int d^{3}X_{m,\overline{m}} \Lambda^6 e^{-2S_c} \sqrt{g_{m}}\sqrt{g_{\overline{m}}} \exp\le[-\frac{i}{2}q_m\le(\th(\Xm) - \th(\Xmb)\ri) - S_{berry} - S_{\p}\ri] \ . \label{sad1}
\ee
For notational convenience we will not write down the partition function in the denominator $Z[0]$; as usual its effect is to remove disconnected diagrams. Thus essentially we evaluate the field on the locations of the monopoles and integrate their locations throughout the bulk spacetime. We now need to determine $\th(X)$. This is the solution of \eqref{chisources} subject to the appropriate boundary condition and can be written formally as
\be
\th(X) = \frac{i q_m}{g_F^2}\le(G(X,\Xm) - G(X,\Xmb)\ri) + \int d^2y' K(X;y')H(y') \ . \label{formsol}
\ee
Here $G(X,X')$ is the bulk-to-bulk propagator \eqref{bulkbulkdef} constructed in the previous section. $K(X;y')$ is the bulk-to-boundary propagator that satisfies the boundary conditions \eqref{bcsource}. 

Now we put this into \eqref{sad1}. There are several terms. We first focus on the terms coming from the bulk-to-bulk propagators, which are
\be
S_{bb} = \frac{q_m^2}{2g_F^2}\le(2 G(\Xm,\Xmb) - G(\Xm,\Xm) - G(\Xmb,\Xmb)\ri) \label{twobb}
 \ee
 Here the first term is a long-range interaction between the two monopoles. Each of the second two terms is clearly singular, corresponding to the infinite self-energy of a point charge. One might have thought this self-energy would be included in the single-monopole action $S_c$; however the dependence of these singular functions on the holographic coordinate $r$ contains extra information, which can be thought of as the energy of the interaction of the monopole field with the nontrivial geometry and boundary conditions. In the calculation it appears as a contribution to the monopole fugacity, which now depends on $r$. We discuss this in the next subsection, and for now we simply call it $S_{self}(r)$, making the full action of the monopole $S_m(r) \equiv S_c + S_{self}(r)$. 

We now turn to the boundary term $S_{\p}$. It is shown in the Appendix in \eqref{bdycont} that on the monopole solution this takes the form
\be
S_{\p} =  \ha \int_{\p} d^2y\;H(y)\le(iq_m (K(\Xm;y) - K(\Xmb;y)) + g_F^2\int d^{2}y' K(r = r_{\Lam},y';y)H(y')\ri) \ . \label{bdycontm}
\ee
The last term knows nothing of the monopoles; it is essentially the tree-level answer, and if we evaluate it on a monopole configuration it results in a disconnected diagram that should be dropped. However the other terms must be included. 

\begin{figure}[h]
\begin{center}
\includegraphics[scale=0.7]{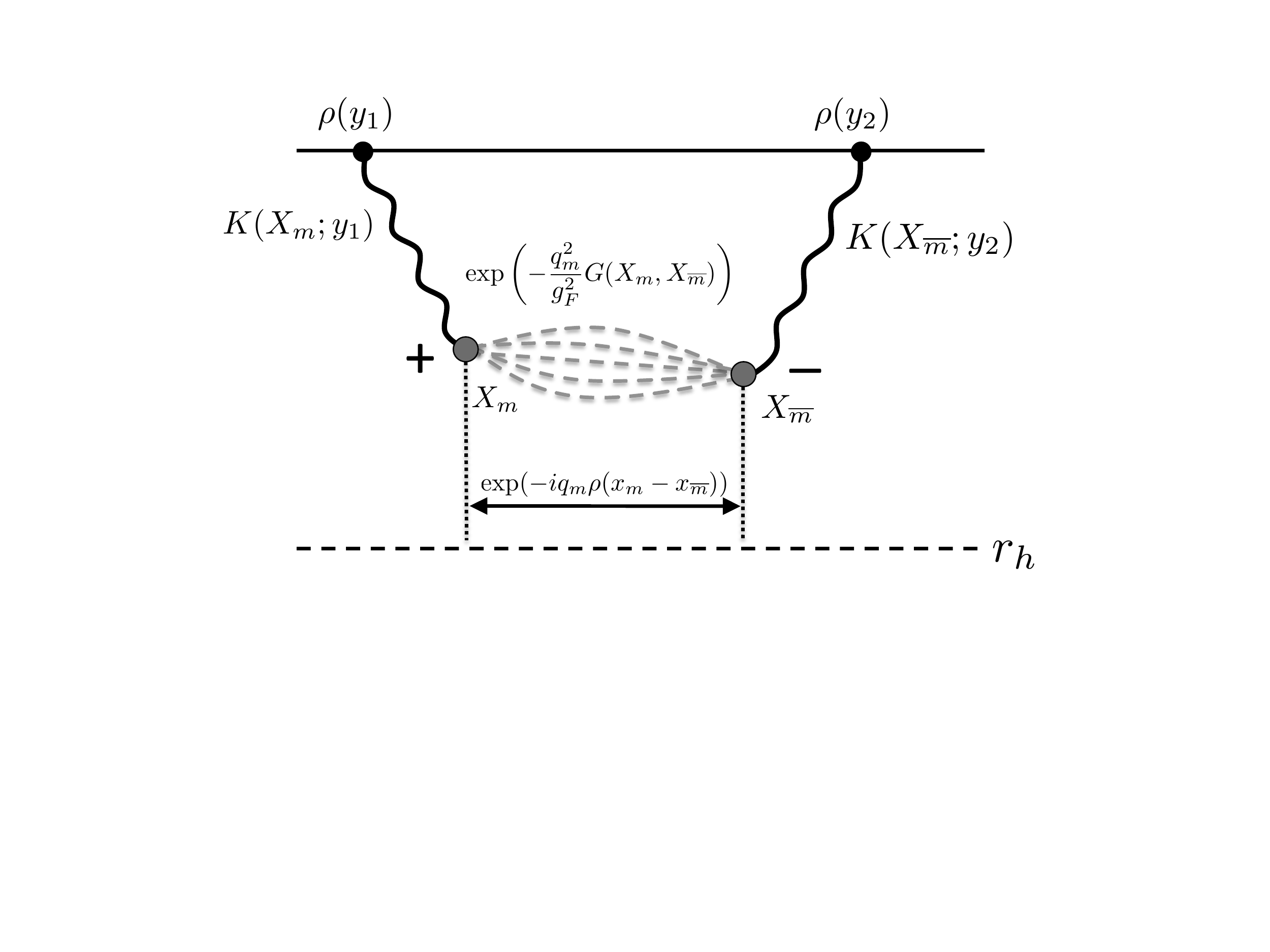}
\end{center}
\vskip -0.5cm
\caption{Witten diagram illustrating one of four terms (the one we compute) in \eqref{wittenans}.}
\label{fig:witten2}
\end{figure}

\begin{figure}[h]
\begin{center}
\includegraphics[scale=0.6]{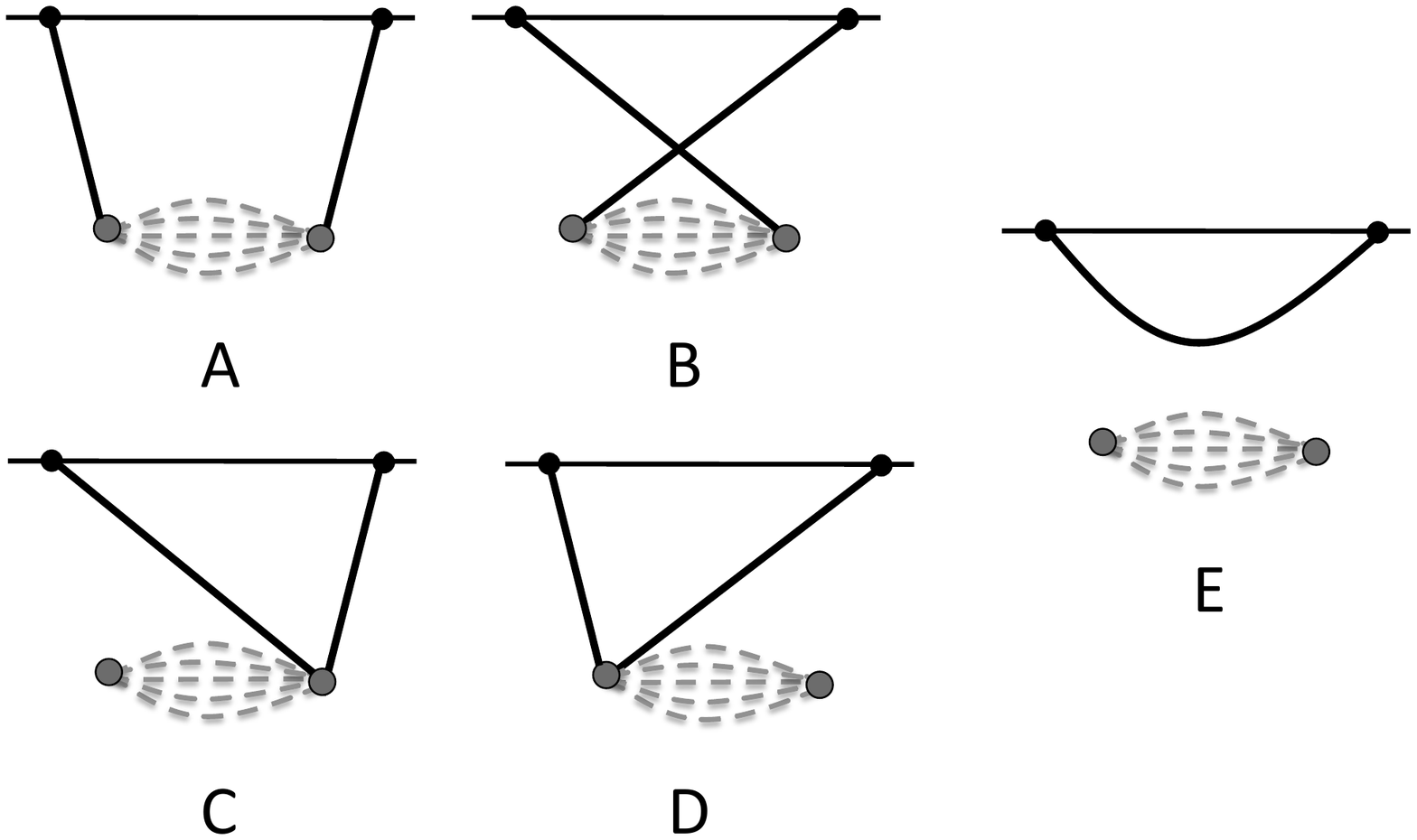}
\end{center}
\vskip -0.5cm
\caption{Different contributions to final answer \eqref{wittenans}, corresponding to different ways to attach boundary points to the bulk monopole. We only compute $A$; $B$ is related to it by switching $y_1$ and $y_2$. $C$ and $D$ are not expected to contribute interesting correlations. $E$ is a disconnected diagram that can be seen in \eqref{bdycontm}: it is essentially equivalent to the tree-level answer and will cancel if we compute a properly normalized two-point function. }
\label{fig:manywittens}
\end{figure}

Collecting all of these terms we see that the full quantity being exponentiated is
 \begin{align}
\exp\bigg[ - S_{m}(\rrm) - S_{m}(\rmb)- & i\rho q_m(\xm - \xmb) - \frac{q_m^2}{g_F^2}G(\Xm,\Xmb) \nonumber \\ & - iq_m\int d^2 y'\le(K(\Xm;y') - K(\Xmb;y')\ri)H(y)\bigg]
\end{align}
Note that in the last term half of the contribution comes from $S_{\p}$ and half from the contribution localized on the monopoles in \eqref{sonsh}. Putting this into \eqref{sad1} and taking the functional derivatives we find the following expression for the boundary theory Green's function:
\begin{align}
\sG(y_1,y_2) = -q_m^2\int d^3X_{m,\overline{m}} \sqrt{g_m} \xi_m\;\sqrt{g_m} \xi_{\overline{m}} & (K(\Xm;y_1) - K(\Xmb;y_1))(K(\Xm;y_2) - K(\Xmb;y_2)) \times \nonumber \\
& \exp\le(-i\rho q_m(\xm - \xmb) - \frac{q_m^2}{g_F^2}G(\Xm,\Xmb)\ri) \label{wittenans}
\end{align}
where we have defined the radially dependent monopole fugacity $\xi_i \equiv \Lambda^3 e^{-S_m (r_i)}$. Each term in this expression can be represented as a Witten diagram as in Figure \ref{fig:witten2}; the two boundary insertions are connected by bulk-to-boundary propagators to monopoles, and these monopoles are then integrated over the bulk with an exponentially suppressed weight that takes into account their interaction. 
There are four terms in the expression above, corresponding to different ways to attach the boundary points to the monopoles, as shown in Figure \ref{fig:manywittens}. We expect nontrivial correlations only when the two boundary points are attached to different monopoles. We now compute one such diagram (e.g. diagram A), denoting it by $\sG(y_1,y_2)_A$. The other interesting diagram (diagram B) is obtained by switching $y_{1}$ and $y_2$. 

It is convenient to switch to momentum space $p = (\om_E, k)$ in the boundary coordinates $y_{1,2}$, after which we find
\begin{align}
\sG(p^1,p^2)_A = -q_m^2 \int d^{3}X_{m,\overline{m}} \sqrt{g_{m,\overline{m}}}& \xi'_{m,\overline{m}}f_{p^1}(\rrm) f_{p^2}(\rmb) \exp\bigg(ip^1_a \ym^a + ip^2_a \ymb^a \nonumber 
 \\ & - i\rho q_m(\xm - \xmb)  - \frac{q_m^2}{g_F^2}G(\Xm,\Xmb)\bigg),
\end{align}
where the $f$'s are the mode expansions of the bulk-to-boundary propagators, which should be understood as connecting a boundary excitation with momentum $p$ to a bulk point $(r,y^a)$:
\be
K(r,y^a;p) = \exp\le(i p_a y^a\ri)f_p(r)
\ee
We now perform the bulk integrals over the monopole locations. It is convenient to define a relative and a center of mass coordinate in the field theory directions, i.e.
\be
\Delta^a \equiv y_m^a - y_{\overline{m}}^a \qquad \Sigma^a \equiv \frac{\ym^a + \ymb^a}{2}
\ee
The integral over $\Sigma^a$ can now be done explicitly to give a delta function, as there is no dependence on the center of mass coordinate. Thus the answer is now $ \sG(p^1,p^2)_A = (2\pi)^2 \delta^{(2)}(p^1 + p^2)\sG(p_1)$, where 
\be
\sG(p) = -q_m^2\int dr_{m,\overline{m}} \sqrt{g_{m,\overline{m}}}\xi'_{m,\overline{m}} d^2\Delta f_{p}(r_m) f_{-p}(r_{\overline{m}}) \exp\le[i p_a \Delta^a - i \rho q_m \Delta_{\x} - \frac{q_m^2}{g_F^2}G(\rrm,\rmb;\Delta)\ri] \label{corr1}
\ee
This is as far as we can go in the abstract. To proceed we need an explicit expression for the potential energy $G(r_1,r_2;\Delta)$ of the two monopoles. This depends on the detailed geometry (and indeed, in some cases, on the {\it fluctuations} of the geometry sourced by the monopoles), and so we will treat the high and zero temperature cases separately. 

\section{High temperature} \label{sec:highT}
We first consider the case when the temperature is very large compared to energy scale associated with the charge density, $\frac{T}{\hat{\rho}} \to \infty$. As discussed earlier this is a probe limit: the charge density does not strongly affect the dynamics, and perturbations of the gauge field decouple from those of the metric. This greatly simplifies the calculation: we are now dealing only with a fluctuating scalar on a fixed background. Thus we take the background to be the $\hat{\rho} \to 0$ limit of \eqref{chbtz}, 
\be
ds^2 = \le(r^2 - r_h^2\ri) d\tau^2 + r^2 d\x^2 + \frac{dr^2}{r^2 - r_h^2} \qquad T = \frac{r_h}{2\pi} \qquad \Th_0 = \rho \x
\ee
i.e. the ordinary uncharged BTZ black hole in planar coordinates. Note that $\p_{\x}\Th \neq 0$ and so we are still working at a nonzero charge density: it simply does not backreact on the metric in the high temperature limit in which we are working. We will often work in Fourier space for the perturbations with $\th(r,x,\tau) = \th(r)e^{i\om_E\tau + i k \x}$, where $\om_E = 2\pi T n$ with $n \in \mathbb{Z}$ are the Matsubara frequencies. 

\subsection{Single monopole field}
\begin{figure}[h]
\begin{center}
\includegraphics[scale=0.45]{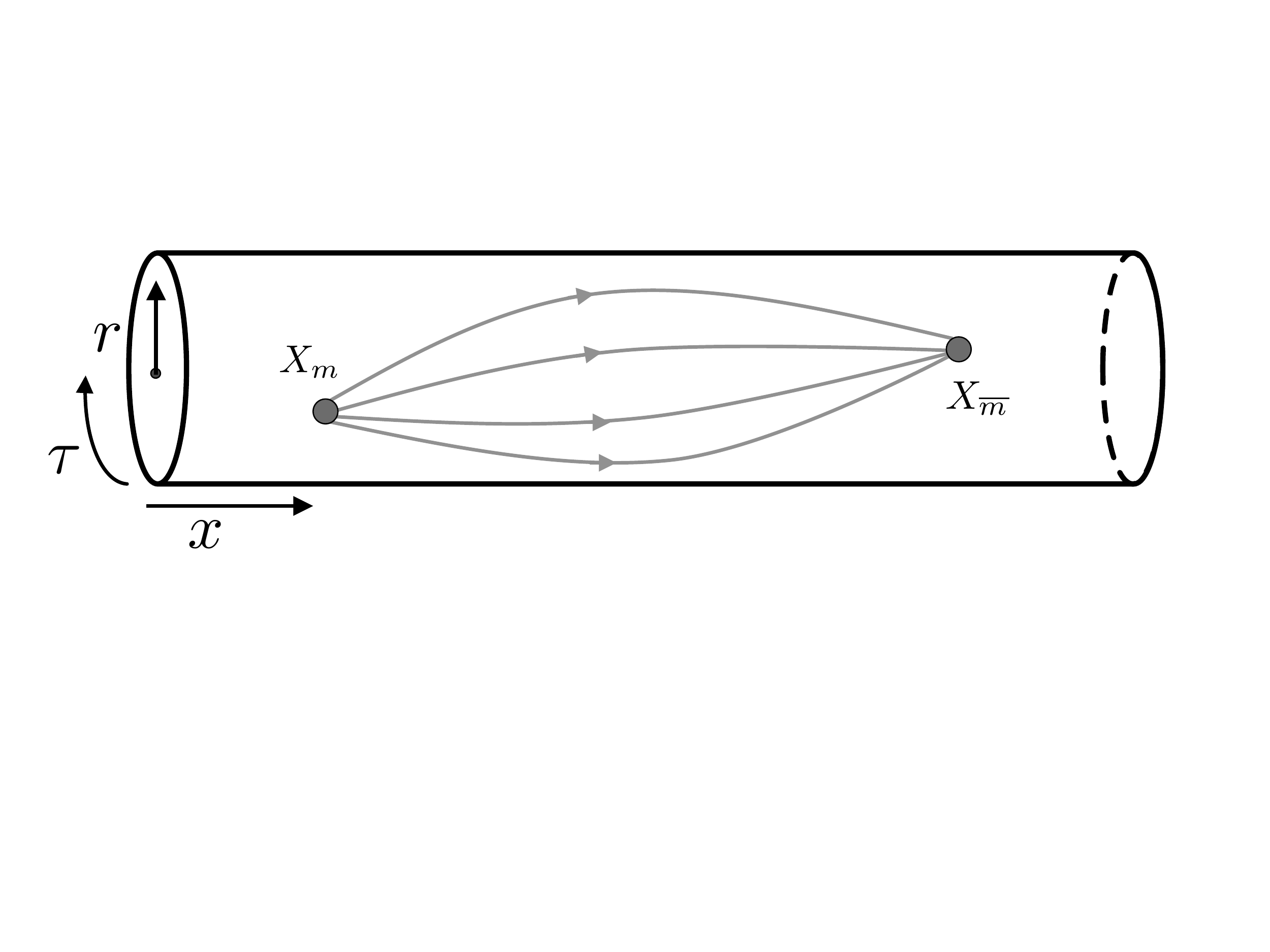}
\end{center}
\vskip -0.5cm
\caption{Monopole-antimonopole pair at large spatial separation and finite temperature. Field lines cannot spread out in $\tau$ (due to compactness) or in $r$ (due to AdS kinematics), making the problem essentially one dimensional.}
\label{fig:highT}
\end{figure}

We first solve for the field of a single monopole on this background. We seek a solution to the equation \eqref{bulkbulkdef}, reproduced here:
\be
\nabla^2 G(X,X') = \frac{1}{\sqrt{g}}\delta^{(3)}(X-X') \ . \label{bulkbulkdef2}
\ee
where the field $\th(X)$ due to a monopole at point $X_m$ is then $\th_{X_m}(X) = \frac{i q_m}{g_F^2}G(X,X_m)$. This is just the usual equation for the bulk-bulk Euclidean propagator, and we can use standard techniques to solve it. In momentum space the propagator can be expressed in terms of solutions to the homogenous wave equation:
\be
\label{ccw}
G(r,r';i\om_E,k) = \frac{\th_b(r_>)\th_{i}(r_<)}{W[\th_b,\th_{i}]}
\ee
with $\th_b$ the solution that is normalizable at infinity, $\th_{i}$ the solution that is regular at the Euclidean horizon, $r_>$ and $r_<$ respectively the larger and smaller of $r$ and $r'$, and $W$ the Wronskian: 
\be
W[a,b] = \sqrt{g}g^{rr}((\p_r a) b - (\p_r b)a)
\ee
We now construct the solutions $\th_b$ and $\th_{i}$. The explicit homogenous wave equation is
\be
\p_r(r(r^2 - r_h^2)\th'(r)) - \le(\frac{r}{r^2 - r_h^2}\om_E^2 + \frac{k^2}{r}\ri)\th(r) = 0
\ee
As the BTZ metric is locally the same as AdS$_3$, we have enough symmetries to solve the wave equation explicitly for all frequency and momentum. The solution that is regular at the Euclidean horizon is
\be
\th_{i}(r) = {_2}F_1\le(\frac{\om_E - i k}{4 \pi T},\frac{\om_E + i k}{4\pi T};1 + \frac{\om_E}{2\pi T};1-\frac{r_h^2}{r^2}\ri)\le(1 - \frac{r_h^2}{r^2}\ri)^{\frac{\om_E}{4 \pi T}}, \label{exactsol}
\ee
where $_2F_1$ is the hypergeometric function. Now the Wronskian is independent of $r$ and so can be evaluated anywhere. A convenient place to evaluate it is at $r = r_{\Lam}$, where we use the fact that $\th_b$ satisfies the boundary condition \eqref{bc1} to find
\be
W = \th_b\le(\Pi_i - \frac{p^2}{\ka}\th_i\ri)\bigg|_{r_{\Lam}} \qquad p^2 = \om_E^2 + k^2 \ .
\ee
Using \eqref{exactsol} it is possible to evaluate this Wronskian for all $\om$ and $k$. However we will eventually be interested in the monopole field at large distances, which will be dominated by the $\om_E = 0$ Matsubara mode and small $k$. In this limit if we normalize $\th_i(r \to \infty) = 1$ at infinity, the Wronskian simplifies to
\be
W(i\om_E = 0, k \to 0) = \th_b(r_{\Lam}) k^2 \log\le(\frac{2\pi T}{r_\star}\ri)
\ee
Note that as claimed $r_{\Lam}$ and $\ka$ have combined into the RG-invariant combination $r_\star = r_{\Lam}e^{\frac{1}{\ka}}$. The fact that the Wronskian vanishes as $k \to 0$ means that there is a pole in the propagator and hence a a long range potential between monopoles separated in the $\x$ direction. To determine the coefficient of the pole we note that in the limit $\om_E = 0$ and $k \to 0$ both $\th_i(r)$ and $\th_b(r)$ coincide and become constant, meaning that we have simply
\be
G(r,r';i\om_E = 0,k \to 0) \sim \frac{1}{k^2 \log\le(\frac{2\pi T}{r_{\star}}\ri)}
\ee
In position space we then have 
\be
G(r,r';\tau,\x \to \infty) = T\sum_{i\om_E}\int \frac{dk}{2\pi} G(r,r';i\om_E,k)e^{i\om_E \tau + i k \x} \sim \frac{T|\x|}{2 \log\le(\frac{r_{\star}}{2\pi T}\ri)} \label{monsep}
\ee
where the last expression correctly captures the leading dependence at large $\x$. We note that the problem has effectively become one-dimensional, as in Figure \ref{fig:highT}. This should not be surprising; essentially the field lines cannot spread out along the Euclidean time circle as it is compact, and the boundary conditions imposed by the AdS kinematics do not allow them to spread out in the AdS radial direction either. Thus they can only extend in the $\x$ direction, and the linearly growing potential in $\x$ is standard for a point charge living in one flat dimension.

Note also that the sign of the potential depends on whether $r_\star$ is an IR scale or a UV scale. If it is far larger than the temperature -- as we have been assuming -- then the overall sign is positive and two monopoles of opposite charge will feel an attractive force, which is physically reasonable.

\subsection{Friedel oscillations at high temperature} \label{sec:highTFr}

Recall that at the end of the previous section we had established the form of the instanton correction to the current-current correlator to be \eqref{corr1}:
\be
\sG(p) = -q_m^2\int dr_{m,\overline{m}} \sqrt{g_{m,\overline{m}}}\xi'_{m,\overline{m}} d^2\Delta f_{p}(r_m) f_{-p}(r_{\overline{m}}) \exp\le[i p_a \Delta^a - i \rho q_m \Delta_{\x} - \frac{q_m^2}{g_F^2}G(\rrm,\rmb;\Delta)\ri] \label{corr1b}
\ee
With the above expression for the monopole-monopole correlation we are now in a position to finish the computation of this correction. We expect the answer to be dominated by the contribution from monopoles that are widely separated in the $\x$ direction. In this regime we can directly use \eqref{monsep}, $G(r,r';\Delta_{\tau},\Delta_{\x} \to \infty) = \frac{T}{2 \log\le(\frac{r_{\star}}{2 \pi T}\ri)} |\Delta_x|$.

At this point we also set the external frequency $\om_E = 0$, keeping only the spatial momentum $k$. In this limit at large $\Delta_\x$ nothing depends on $\Delta_\tau$, and so we can explicitly do the integral, picking up a factor of the inverse temperature. The remaining integral over $\Delta_\x$ is:
\be
I(k) \equiv \frac{1}{T}\int d{\Delta_\x} \exp\le(i k \Delta_\x - i \rho q_m \Delta_\x - 2\pi T K^T |\Delta_\x|\ri) \qquad K^T = \frac{q_m^2}{g_F^2}\frac{1}{4\pi \log\le(\frac{r_{\star}}{2 \pi T}\ri)}
\ee
where the definition of the parameter $K^T$ is for later convenience. This expression should be understood as correctly capturing the contribution to the integral from large $\Delta_\x$. The integral is a Lorentzian and we find
\be
I(k) = 4 \pi K^T  \frac{1}{\le(2 \pi T K^T\ri)^2 + \le(k - \rho q_m\ri)^2} \label{ikexp}
\ee
This is the interesting momentum-dependence of the final answer. The peak of the Lorentzian is at a $k$ value that is shifted by $\rho q_m$; this is precisely the desired Friedel oscillation, and the wavevector of the oscillation could clearly have been predicted from the Berry phase term alone, with essentially no computation. However the detailed form of the finite-temperature Lorentzian followed from detailed considerations involving the bulk geometry. We leave further discussion of the result to the conclusion section.

We now note that there still remains a momentum-dependent radial integral to do. The full answer takes the form
\be
\sG(k) = -q_m^2 \Lam^6 R(k)^2 I(k)
\ee
where the radial integral has been absorbed into a function,
\be
R(k) = \int^{r_{\Lam}} dr \sqrt{g} e^{-S_m(r)}f_k(r) \label{rkint}
\ee
and we have used the fact that $R(k) = R(-k)$. This expression determines the effective contribution of a monopole at radius $r$ to the answer. In the field theory each monopole corresponds to some sort of instanton event, with its radial position $r$ loosely corresponding to the size of the corresponding instanton. Thus this expression can be interpreted from the field theory point of view as some sort of monopole ``density of states''.

$f_k(r)$ is the mode expansion of the bulk-to-boundary propagator, and so is simply the (appropriately normalized) solution to the wave equation that is regular at the horizon \eqref{exactsol}. We are interested in this expression at very large momentum, $k \sim q_m \rho \sim N^2$. Note that this very large Euclidean momentum means that we are very off-shell, and we expect the mode function to fall off quickly as we penetrate to the bulk of the spacetime. While this can be confirmed from examining \eqref{exactsol}, it is simpler to realize that at high momentum we are probing the high-energy CFT$_2$ structure of the theory, and thus (at least at $r \gg r_h$) the mode function should be the appropriate solution to the wave function on pure AdS$_3$ in the Poincare patch, i.e. the Bessel function
\be
\label{f}
f_{k \to \infty}(r \gg r_h) \sim \frac{K_1\le(\frac{k}{r}\ri)}{r k \log\le(\frac{k}{\bar{r}_{\star}}\ri)} \ . \qquad \bar{r}_{\star} = 2 r_{\star} e^{-\ga}
\ee
Here the normalization has been fixed by requiring that at infinity $f_p(r)$ corresponds via \eqref{bcsource} to a source of unit magnitude for all momentum: $H(k) = 1$. Note that
in order to avoid problems with the UV pole at $k= \bar{r}_\star$ we must demand that
$ k \sim q_m \rho \ll r_\star$. This is easy to achieve with a moderately small $\ka \sim 1/\log(N)$.

Note now that at intermediate radii $r_h \ll r \ll k$ we have from the expansion of the Bessel function $f_k(r) \sim \exp\le(-\frac{k}{r}\ri)$, and thus the mode function is indeed exponentially suppressed in the bulk. This means that (modulo contributions from $S_{self}(r)$) the bulk of the contribution comes from monopoles very close to the AdS boundary; the corresponding field-theoretical instantons are almost as small as can be, as the large ones do not couple strongly to the hard probe that is a high-momentum correlation function 

There is also some radial dependence in the function $S_m(r)$. As explained in detail in Appendix \ref{sec:fug}, this corresponds to a radially dependent monopole fugacity. While it can in principle be computed in terms of a sum over the mode functions, this is numerically very time-consuming for the BTZ black hole. We instead compute the fugacity for pure AdS$_3$, as the dominant contribution to the monopoles comes from radial positions $r \gg r_h$ and we expect this to be a good approximation to the answer. It is
\be
S_m(r) = S_{core} -\frac{q_m^2}{4\pi g_F^2}\log\log\le(\frac{\bar{r}_{\star}}{r}\ri),
\ee
This result is discussed further in the Appendix. It wants the monopoles to be towards the interior of the geometry, but clearly depends very weakly on $r$ and arises purely from the marginally running boundary conditions. 

Thus the gentle dependence on $r$ of the fugacity does not change the fact that the integral in \eqref{rkint} is UV-divergent. There is a cutoff-dependent piece in the answer:
\be
R(k) \sim \le(\frac{r_{\Lam}}{k}\ri)^2 e^{-S_{m}(r_{\Lam})}\log^{-1}\le(\frac{k}{\bar{r}_{\star}}\ri) \label{rkfin}
\ee
We may thus assemble all of the pieces to obtain the full form of the Friedel oscillation at finite temperature:
\be
\langle \delta \rho(k) \delta \rho(-k) \rangle =  e^{-2S_m(r_{\Lam})}\frac{A}{\le(2 \pi T K^T\ri)^2 + \le(k - \rho q_m\ri)^2} + (k \to -k) \label{finansT}
\ee
where we have used \eqref{currcurr} to relate the answer to a density-density correlation and the overall constant $A \sim 4\pi g_F^{-4}\Lam^6 K^T\frac{r_{\Lam}^4}{\rho^2} \log^{-2}\le(\frac{q_m \rho}{\bar{r}_{\star}}\ri)$. We have also added in the contribution from the other contributing diagram (i.e. diagram B) in Figure \ref{fig:manywittens}; this results in an answer that is symmetric in $k$ by creating the corresponding singularity at $k = -\rho q_m$. In determining the prefactor we have set $k \to \rho q_m$, as this expression should be understood only as capturing the leading momentum dependence in the vicinity of the $\rho q_m$. 

\section{Zero temperature} \label{sec:zeroT}

We turn now to the structure of the Friedel oscillations
at zero temperature. These oscillations will occur at the same wavevector -- $\rho q_m$ -- as the
large temperature analysis of the previous section. However the form of the singularity should be different. In particular in momentum space the Lorentzian \eqref{finansT}, whose
width vanishes as $T \rightarrow 0$, should become a genuine singularity whose form we would like to determine.

\subsection{Sound mode}

While our final goal is to compute the contribution of monopoles
to the density-density correlator at zero temperature in the charged BTZ black hole, we
start by computing the answer in the absence of the monopoles.
We go through this in some detail, both because it is interesting in its own right and we
will use the results heavily later in determining the form of the long-range interaction between monopoles. Interestingly we will find a zero temperature
sound mode similar to the sound mode found in the higher dimensional
charged black hole \cite{Edalati:2010pn}.  This sound mode will turn out to control the form
of the Friedel oscillations. 

Since the charged black hole mixes the scalar field (or the gauge field in the dual
representation) with the metric the dynamical problem becomes much
more difficult. This sort of problem has been addressed in a similar
context by \cite{Edalati:2010pn,Edalati:2010hk}
and we follow their methods. The key is to find
master variables that diagonalize this mixing. We will call these
modes ``gauge invariant'' for reasons which will be clear.

\subsubsection{The gauge invariant variables}

We start by perturbing the charged BTZ background (\ref{chbtz}) using the
dual scalar variable $\Th$. 
We pick a gauge for the metric fluctuations 
suited for holographic computations $g_{ra} = 0$ and $g_{rr} = f(r)^{-1}$:
\bea
ds^2 &=& f(r)\left( 1+  \htt(X) \vphantom{\hxx(X) } \right) d\tau^2 + r^2 \left( 1 + \hxx(X) \right) d\x^2  
+ 2 r^2 \hxt(X) d\tau d\x + \frac{dr^2}{f(r)}\,,  \\
\label{chir}
 && \hspace{3cm} \Theta  =  \left( \frac{1}{16 \pi G_N g_F^2} \right)^{1/2}  \left( \hrho \x +  \dchi (X) \right) \,. 
\eea
Note we have rescaled $\Th = \left( 16 \pi G_N g_F^2 \right)^{-1/2} \hat{\Th}$
for future convenience. 
The same rescaling of $\rho \rightarrow \hat{\rho}$ was introduced in Section~\ref{chbtzsec}.
The linearization of the Einstein equations and the scalar equation for $\Th$ defines the dynamical
problem we would like to solve. The equations will be left in Appendix~\ref{back} since they are not very enlightening.

It is useful to note that there are three unfixed diffeomorphisms $\{ r \rightarrow r + \sqrt{f} \zeta^r \,, \x \rightarrow \x + \zeta^\x ,\, \tau \rightarrow \tau + \zeta^\tau \}$  under which these fields transform:
\bea
\nonumber
\dchi & \rightarrow& \dchi + \hrho \zeta^\x \,, \hspace{3.5cm} \hxt \rightarrow \hxt + \partial_\tau \zeta^\x + \frac{f}{r^2} \partial_\x \zeta^\tau \,,  \\
\label{allg}
\htt &\rightarrow& \htt + 2 \partial_\tau \zeta^\tau + \frac{f'}{\sqrt{f}} \zeta^r   \,,  \qquad \hspace{.5cm}
\hxx \rightarrow \hxx + 2 \partial_\x \zeta^\x + \frac{2\sqrt{f}}{r} \zeta^r \,,
\eea
where the $(\zeta^r,\zeta^\x,\zeta^\tau) $'s satisfy the following :
\be
\partial_r \zeta^r = 0\,, \qquad \partial_r \zeta^\x = - \frac{1}{r^2 \sqrt{f}} \partial_\x \zeta^r\,,
\qquad \partial_r \zeta^\tau =- \frac{1}{f^{3/2}} \partial_\tau \zeta^r \, .
\ee
There are $4$ independent fields $(\dchi, \hxt, \htt, \hxx)$ each with second order equations of motion
in the radial direction. Together with the 3 first order constraints from Einsteins equations
and 3 unfixed diffeomorphisms  we should be left with two independent first order degrees
of freedom (or one second order). 
This is of course expected since in three dimensions gravity does
not contribute any propagating modes. In order to cleanly solve this problem
we should write down two degrees of freedom which are gauge invariant
under the left over diffeomorphisms (\ref{allg}). We choose these as:
\be
\label{ginv}
\sigma^1 =  D_1 \dchi -  \hrho f \partial_\x \htt + \frac{\hrho f' r}{2}  \partial_\x \hxx + 2 \hat{\rho} r^2 \partial_\tau \hxt \,,
\qquad
\sigma^2 = 2 r f \dchi' - 2 \partial_\x^2 \dchi + \hrho \partial_\x \hxx\,,
\ee
where $D_1 \equiv - \left( f' r \partial_\x^2 + 2r^2 \partial_\tau^2 \right)$.
From now on we will work in Fourier space with our conventions
given in \eqref{fconv}.
The dynamical equation for the $\sigma$'s may then be derived from the
equation of motion (\ref{dyn1}-\ref{dyn4}) and constraints (\ref{const1}-\ref{const3}) by brute force. 
See Appendix~\ref{back} for a general discussion of this.
They must close on themselves because we know they are gauge invariant and we know
there are only two independent such variables:
\be
\label{eqsigma}
 \partial_r \begin{pmatrix} \sigma^1 \\ \sigma^2 \end{pmatrix} = 
\frac{1}{fr} \begin{pmatrix} 2 f - k^2 &  D_1/2+  \hrho^2 f 
\\ - \frac{2(k^4 -  \omega_E^2 r^2 - k^2 f)}{D_1} & \frac{ \hrho^2 k^2 f}{D_1} + k^2 \end{pmatrix} \begin{pmatrix} \sigma^1 \\ \sigma^2 \end{pmatrix}  \,.
\ee

Start by imposing a standard regularity condition on the $\sigma$'s in the interior
at $r=r_h$. For example at $T=0$ this solution behaves as follows close to the horizon:
\be
\begin{pmatrix} \sigma^1 \\ \sigma^2 \end{pmatrix} \sim \begin{pmatrix} \hrho^2 \omega_E^2 \\ 4 k^2 + 2 |\omega_E| \hrho \end{pmatrix} (r-r_h)^{\frac{ |\omega_E|}{3 \hrho}} e^{ - \frac{ | \omega_E| }{2(r -r_h)} } \, .
\ee
 Solving \eqref{eqsigma} at large $r$ we find the following behavior:
\be
\label{asigma}
\sigma^1 = \alpha r^2 + \ldots \,, \qquad \sigma^2 = \alpha \log r + \beta + \ldots \,,
\ee
where the ratio $\beta/\alpha$ will be fixed by the regularity condition.
We impose the following boundary conditions on the metric:
\be
\label{bmetric}
\hxx \rightarrow 0 + \mathcal{O}(r^{-2}) \,,\qquad
\htt \rightarrow 0 + \mathcal{O}(r^{-2}) \,,\qquad
\hxt \rightarrow 0 + \mathcal{O}(r^{-2})  \,,
\ee
since for now we will not be interest in sourcing the stress tensor in the field theory. 
These boundary conditions are enough to fix the diffeomorphisms given in (\ref{allg})
and thus we can in principle solve for all the dynamical fields subject to the constraints
from Einsteins equations. So we now have enough information to extract the boundary behavior of $\dchi$ using (\ref{ginv}) and (\ref{asigma}-\ref{bmetric}) we find,

\be
\label{largechi}
 \dchi = \frac{\alpha}{2p^2} \left( 1 + \frac{p^2}{2r^2} ( k^2/p^2 - (\beta/\alpha) -1/2 - \log r )  + \ldots \right) \,.
\ee
This can then be used to extract the current current function (after
subtracting the background density) using \eqref{jjcorr}.
\be
\label{fincorr}
\left< \delta j^a \delta j^b \right> =  \frac{1}{g_F^2} \frac{ \epsilon^{ac} \epsilon^{bd} p_c p_d}{
k^2 - p^2(\beta/\alpha + \log(r_\star) ) }
\ee
Note that it is trivial to analytically continue this Euclidean answer to real times $\omega_E
\rightarrow - i \omega$. See
\cite{Iqbal:2009fd} for a discussion of this. 
In the higher dimensional case the same problem has been solved \cite{Edalati:2010pn}
and an examination of the current-current correlator studied in \cite{Edalati:2010pn, Hartnoll:2012wm,Davison:2011uk} reveal no $2 k_F$-like singularities. This should continue to hold
in the lower dimensional case at hand, despite certain details being different.
We have done some unenlightening numerical calculations of \eqref{fincorr} which
seem to confirm this suspicion. 


\subsubsection{Hydrodynamic limit}
\label{hydlim}

We can miraculously solve \eqref{eqsigma} in the small $p$ limit.
This is analogous to the usual finite temperature hydrodynamic limit, where instead
of requiring the momenta to be small in units of the temperature we
require they are small in units of $\hat{\rho}$.  
This allows us to extract the zero temperature sound mode. 
For $\omega_E^2, k^2 \ll \hrho^2$ we have,
\be
\label{hydeq}
 \partial_r \begin{pmatrix} \sigma^1 \\ \sigma^2 \end{pmatrix} = 
 \begin{pmatrix} 2/r  &   \hrho^2/r
\\  \frac{2(\omega_E^2 r^2 + k^2 f )}{f D_1} & \frac{ \hrho^2 k^2 }{D_1} \end{pmatrix} \begin{pmatrix} \sigma^1 \\ \sigma^2 \end{pmatrix} \,,
\ee
which has a solution
\be
\label{hydsoln}
\sigma_1 = \alpha f \qquad \sigma_2 = \frac{\alpha}{ \hrho^2} \left( r f' - 2 f \right) \ .
\ee
We found this solution with the same methods used to
construct Goldstone modes of holographically ordered phases 
\cite{Iqbal:2010eh}. This solution is regular in the interior and so we
can use it to read off: $\beta/\alpha = -\log(r_h) + r_h^2/\hrho^2 - 1/2$.
The density correlator in $p$ space (\ref{fincorr}) in the hydro limit is:
\be
\label{hydjt}
\left< \delta \rho(p) \delta  \rho  (-p)\right> = \frac{1}{g_F^2} \frac{Z k^2}{\omega_E^2 + v_s^2 k^2+ \Sigma} \,,
\ee
where,
\be
 \quad Z = \frac{1}{ \log(r_\star/r_h) + 1/2 - r_h^2/\hrho^2} \,,\quad\, v_s^2 = 1 - Z\,.
 \ee
Note that $\delta j^{\tau} = i\delta\rho$. Clearly there is a sound mode - in real time there is a mode which
disperses as $\omega = \pm v_s k$. To make sure this is a genuine sound mode we
have to show that the width from the self energy $\Sigma$ is small in the hydrodynamic limit.
This is computed in Appendix \ref{hydroapp}  where we find the
width is highly suppressed $\Sigma(\omega_E \rightarrow i \omega) \sim i \omega^5$.

At zero temperature the speed of sound satisfies:
\be
v_s^2 = 1 - \frac{1}{\log(r_\star/r_h)} = \frac{\rho}{\mu} \frac{d \mu}{d\rho} \label{vs}
\ee
where we have used the thermodynamics of the charged BTZ black hole
discussed around \eqref{susc}. 
As discussed in Section~\ref{chbtzsec} we are only interested in the regime where $r_h \ll r_\star$
where it is clear that $0 < v_s^2 < 1$.

An analogous sound mode was found 
in a probe brane setup \cite{Karch:2008fa}
where it was suggested to be the holographic realization of the zero sound mode of an
underlying Fermi surface.  Thus giving evidence to the fermionic nature of that phase.
It seems reasonable to make a similar identification
in the charged black hole setup studied here and in higher dimensions \cite{Edalati:2010pn}. 
A more fine grained comparison of this zero sound mode to a Landau Fermi Liquid (LFL) was
undertaken both for a probe brane \cite{Davison:2011ek} and the higher dimensional charged
black holes \cite{Davison:2011uk} with the former behaving more like a LFL. 

The mode we found will be crucial in revealing the form of the $2 k_F$ singularity. 
This is similar to what one would expect for a Luttinger
liquid, where the sound mode is described by a massless scalar $\phi$
and the $2k_F$ singularities are created by operators such
as $e^{ 2 i \phi}$. For further discussion of the relation to a Luttinger liquid see the conclusions.
All in all, we feel this serves as further evidence for the zero-sound interpretation
of these modes.

\subsection{Friedel oscillation computation} 
\label{friedmatch}

We are now in a position to compute the monopole anti-monopole potential
for monopoles situated at large $r \gg r_h$. We concentrate on the monopoles living at large $r$ simply because for smaller $r$ their contribution to the Friedel oscillation computation is highly suppressed by the bulk to boundary propagator.\footnote{Rather 
the appropriate \emph{probes} of the monopoles which necessarily have large
momentum cannot penetrate to smaller $r$. } In Section \ref{sec:highTFr} we have already argued
this to be the case at high temperature, and it remains  the case here since the bulk to boundary propagator at large momentum is insensitive to the density or temperature and will continue to 
be given by \eqref{f}.

At large $r \gg r_h$ the form of the bulk fields in momentum
space is still \eqref{largechi} and \eqref{bmetric} (actually here
we are also taking $r \gg k,\omega_E$). In particular the effects
of mixing between $\dchi$ and the gravitational fluctuations are suppressed.
Thus we can make use of the formula \eqref{ccw} for the bulk-bulk propagator:
\be
\label{neww}
G(r,r';i\om_E,k) = \frac{\hat{\th}_b(r_>)\hat{\th}_{i}(r_<)}{W[\hat{\th}_b,\hat{\th}_{i}]} \,,
\ee
where the regular solution $\hat{\th}_i (r_<)$ is proportional
to that given in \eqref{largechi} and the boundary normalizable solution is:
\be
\hat{\th}_b(r_>) \propto 1 + \left(\frac{p}{2r_>}\right)^2 \left(-1 + 2 \log(r_\star/r_>) \right)  \,.
\ee
For our purposes here we have kept only the dominant terms
at large $r,r' \gg k,\omega_E,r_h$. The  wronskian in \eqref{neww} is
then simply the $AdS_3$ wronskian for the massless scalar wave equation: 
$W[a,b] \approx r^3((\partial_r a) b -(\partial_r b) a)$. 
Plugging into \eqref{neww} and keeping the lowest order terms we derive
the unsurprising expression:
\be
\label{Gmatch}
G(r,r';i \omega_E,k) =\frac{1}{k^2 - p^2(\beta/\alpha + \log(r_\star) ) }\,
\underset{\omega_E , \,k \ll r_h}{=} - \frac{Z}{\omega^2_E + v_s^2 k^2}.
\ee
This is essentially the same  (up to normalization) as the boundary greens function for $\dchi$
since the appropriate bulk wave functions are approximately constant
at large $r$.  For the interaction energy at large separation we only need the small $k,\omega_E$ limit.
Taking the fourier transform, and cutting off the $k,\omega_E$ integrals at small $ v_s^2 k^2 +\omega_E^2  >  1/L_{IR}^2$ we find at large separation,
\be
G(r,r';\tau,x) =  \frac{  Z}{4\pi v_s} \log\left( \frac{\x^2 + v_s^2 \tau^2}{L_{IR}^2} \right) + \ldots \,.
\ee
Note that the IR cutoff $L_{IR}$ will drop out of the final result due to a similar IR divergence
that appears in the momentum integrals that define the monopole self energy, as is discussed further in Appendix \ref{sec:fug}.  

We can now evaluate \eqref{corr1} and integrate over the separation. This gives
us the frequency and momentum dependence close to $k \sim q_m \rho$, i.e. the zero temperature generalization of \eqref{ikexp}:
\be
\label{finI}
I(\omega_E,k) \sim  \left( v_s^2 (k - q_m \rho)^2 + \omega_E^2 \right)^{K} \,,\qquad
K =  \frac{ q_m^2 Z}{4\pi g_F^2 v_s}\,.
\ee
This demonstrates the form of the zero temperature singularity that we were
seeking. We have omitted an $O(1)$ correction to the exponent, as $K$ is very large, scaling like an inverse power of $g_F^2$. The key difference from the finite temperature case is that now the gapless sound mode creating the long-range correlation behaves like a massless scalar in the two field-theory directions, resulting in this power-law structure. The radial dependence of the fugacity at large $r$ will essentially be the same as for the high temperature answer so the radial integrals and determination of the form factor $R(k)$ are precisely as discussed in Section \ref{sec:highTFr}. 


\subsection{Monopoles interacting with 3d gravity} \label{sec:3dgrav}

We have obtained an answer. However the alert reader will have noticed that we did not carefully deal with the interactions of the monopoles with gravity; instead we were fortunate that the dominant part of the answer came from a region where these interactions are suppressed. In other words we never wrote down the generalization
of the monopole partition function \eqref{z11} which allows for fluctuations
of the metric. 

This is for a good reason: the fully backreacted problem is both technically and conceptually difficult, forcing us to confront various issues related to monopoles and background electric fields interacting with
 3d gravity. For this particular problem it turns out that these issues do not affect the answer. The resolution was argued for above: the monopoles
 want to live at large $r$ where the interactions with the metric that caused
 these issues in the first place are suppressed. 
 \footnote{Of course the mixing of $\Th$ with gravity is still important since the
 long range fields probe the full geometry, not just the geometry at large $r$. }
Despite this fact, in this section we still go through the argument in some detail, both because it helps us understand some aspects of the above analysis and because it might be useful in future investigations.
We dedicate this section to understanding various issues of monopoles
interacting with 3d gravity, ending with the generalization of \eqref{z11} and 
the computation which leads to \eqref{finI}. 

\subsubsection{Monopole size}
\label{pertframe}

The first issue we must address is the back reaction of a monopole
on gravity. To do this we start in $3d$ flat space with no electric field.
It is natural to rescale $\Th = (16 \pi G_N g_F^2)^{-1/2} \hat{\Th}$ as was done
in \eqref{chir}. A monopole sitting at the origin has the solution:
\be
\partial^2 \hat{\Th} = i \hat{q}_m \delta^3(X) \quad \rightarrow
\quad \hat{\Th} \sim \frac{i \hat{q}_m}{|X|} \,,
\ee
where we introduced the  parameter $\hat{q}_m$ defined as,
\be
\hat{q}_m = q_m \left( \frac{16 \pi G_N}{g_F^2} \right)^{1/2} \,,
\ee
and which has the units of length. The back reaction on gravity is then roughly:
\be
\partial^2 \delta g \sim T^{\hat \Th} \sim (\partial\hat{\Th})^2 \sim\frac{
\hat{q}^2_m}{|X|^4} \quad \rightarrow \quad \delta g \sim \frac{\hat{q}_m^2}{|X|^2}
\ee
So this is large when $|X| \sim \hat{q}_m$. That is, close to the monopole the perturbative
flat space solution breaks down, the magnetic field backreacts on gravity, and we must find a new solution. Indeed one is available: the wormhole solution of \cite{Gupta:1989bs} has long-range fields corresponding to those of a magnetic monopole in flat space, but is actually a Euclidean wormhole threaded by magnetic flux and connecting two asymptotically flat regions. 
In gravity there is thus a natural core size for these monopoles
of order $\hat{q}_m$ -- this is the size of the throat of the above wormhole. In fact there are several ways to complete the linear solution with a non-linear solution, such as for example the original $SU(2)$ 't Hooft-Polyakov monopole where the 
size of the core of the monopole will be determined by the $W$-boson mass scale $M_W^{-1}$ (as long as this length scale is bigger than $\hat{q}_m$.) 

We do not wish to be specific about the exact form of this bulk UV completion. 
This is possible as long as the core size $M_W^{-1}$ or $\hat{q}_m$ is much smaller
than the $AdS$ radius $L$. We can then argue that the ``linearized'' solution is correct
outside the core and imagine matching onto the non-linear core using the flat space 
answer. Then the only effect of the bulk UV completion is on the action associated
with the core $S_c$ introduced in \eqref{mixed2}. Either way we will
always have the constraint that $\hat{q}_m \ll L$. Recall that we fixed units
by setting $L=1$: we briefly restore it in this section only for illustrative purposes. 

The linearized monopole solution in the charged BTZ black hole is
then constructed perturbatively in $\hat{q}_m / L$,
\bea
\label{pertf}
g &=& g_0 + (\hat{q}_m/L) g_1 + \mathcal{O}(\hat{q}_m/L)^2  \,, \\
\hat{\Th} &=& \hat{\Th}_0 + (\hat{q}_m/L)  \hat{\Th}_1  + \mathcal{O}(\hat{q}_m/L)^2 \,,
\label{pertf2} 
\eea
where $g_0$ and $\hat{\Th}_0 = \hrho x$ are the charged black hole solutions
without monopoles.  This expansion will break down close to the monopole,
but in a controlled way. It is important that when we write this perturbative expansion we 
are working with fields that vary on field theory scales all of order $\left\{(r/L^2),\omega_E,k, \ldots \right\} \sim \hrho$, where we introduced this rescaled density in \eqref{jt} and we reproduce it here:
\be
\hrho = (16 \pi G_N g_F^2)^{1/2} \rho \,.
\ee
The action for the solution including the monopoles will then have an expansion of the form:
\be
\label{sadac}
S_E = \frac{L}{16\pi G_N} \left( \hat{S}_0 + (\hat{q}_m/L) \hat{S}_1 + (\hat{q}_m/L)^2 \hat{S}_2  + \ldots \right)\,.
\ee 
Note that the 3 leading terms in \eqref{sadac} are already evident from the high temperature answer \eqref{sonsh}.
The first term is simply the free energy of the charged black hole
and cancels from the final answer so we ignore it. The second
term is the berry phase term and takes this form when written in terms of $\hrho$,
that is $\hat{S}_1 = i \hrho ( \xm - \xmb)$. 
The last term is the monopole interaction energy and fugacities.  Actually
the only difference between the high temperature calculation
and the zero temperature calculation is the fact that 
at first order the metric is non-zero. At second order the metric will
be non-zero in both cases. So we still secretly required that $\hat{q}_m \ll L$
for the high temperature calculation. 

Writing the action \eqref{sadac} in this way makes it clear we are doing
a Euclidean saddle computation of some
path integral which we will specify shortly. Quantum corrections will be down
by factors of $G_N$. \footnote{Note that various fields are (e.g $\Th_1$ and $g_1$) are purely imaginary
on this saddle. We should not be distressed by this. In our Euclidean calculation the classical field configuration we are evaluating is related to a quantum tunneling amplitude; the interplay between the real and imaginary parts of the action of this configuration carries information related to the imaginary Berry phase and so presumably is related to quantum mechanical interference between different tunneling paths.}

\subsubsection{Monopole position}
\label{sec:monpos}

Naively following the steps in Section~\ref{sec:monsum}
we should be performing a path integral over both $g$ and $\Th$ 
defined by integrating the monopole partition function \eqref{z11} over metrics weighted by the usual Einstein-Hilbert action:
\be
\label{z11g}
Z^{(1,1)} = 
\int [\mathcal{D} g] [\mathcal{D}\hat{ \Th}] \int d^3 \Xm \sqrt{g(\Xm)} \int d^3 \Xmb \sqrt{g(\Xmb)} \exp\left(
-S[\Xm,\Xmb] \right)\,,
\ee
where the action (written in terms of the rescaled variables defined above) is now:
\be
\label{s11g}
S[\Xm,\Xmb] = \frac{1}{16 \pi G_N} \left( \int d^3X \sqrt{g} \left(- R - 2  + \frac{1}{2} (\nabla \hat \Th)^2\right)  + i \hat{q}_m (\hat{\Th}(\Xm) - \hat{\Th}(\Xmb) ) \right) \ .
\ee

Varying this with respect to $g,\Th$, the equations of motion for the saddle are:
\be
\label{chiref}
R_{\mu \nu} - \frac{1}{2} g_{\mu\nu}\left( R + 2 \right) = \ha T_{\mu\nu}^{\hat{\Th}} \, ,
\qquad\sqrt{g} \nabla^2 \hat{\Th} = \sum_{i=m,\overline{m}} \pm i \hat{q}_m  \delta^{(3)}(X-X_i) , 
\ee
where we have also defined the appropriately rescaled scalar stress tensor:
\be
T_{\mu\nu}^{\hat{\Th}} = \le(\nabla_{\mu}\hat{\Th}\nabla_{\nu}\hat{\Th} - \ha g_{\mu\nu}(\nabla \hat{\Th})^2\ri) \ .
\ee
We now come to a distressing point. These equations seemingly lead to an inconsistency since the stress
tensor is not conserved:
\be
\label{noncons}
\nabla^\mu T^{\hat{\Th}}_{\mu\nu} =  \nabla^2 \hat{\Th} \partial_\nu \hat{\Th}
=\smmb \pm i \hat{q}_m \partial_\nu \hat{\Th}(X_i)\,.
\ee 
The right hand side of this expression is not obviously zero. This contradicts Einstein's equations: gravity must always be coupled
to a conserved stress tensor. Upon further reflection, this violation of momentum conservation is expected; it is intimately related to the Berry phase and can be understood in terms of the momentum carried by the monopole event, as is discussed further in Appendix \ref{sec:berry}.
Nevertheless, this is a serious problem since we cannot proceed if we do not have a consistent set of equations to solve. 

The resolution to this puzzle is extremely
simple. We should have also extremized the action \eqref{s11g} with respect to the position
of the monopoles. Said differently we should do the integrals  over
monopole positions in \eqref{z11g} within the saddle approximation - 
thus letting the monopoles roam free 
in the bulk and telling us where they want to sit. Varying with respect to the $X_i$ we find,
\be
\label{monpos}
\partial_\nu \hat{\Th}(X_i) = 0\,, \qquad i = m,\overline{m}\,.
\ee
This solves the contradiction presented in \eqref{noncons}. The situation
is the same as if we had coupled a massive particle to gravity. The world line
of the particle must be a geodesic: otherwise the stress tensor associated
with the world line is not conserved.

Note that the situation is different from the case without backreaction. The position of of a given monopole is no longer a free parameter which we integrate over in the end. Rather it depends on the fields $\hat{\Th}$ and $g$, which in turn depend on the position of the other monopoles.
We can trace the reason for this to the fact that diffeomorphisms $\zeta^\mu$ move around
the position of the monopole insertions $\Theta(X_m) \rightarrow \Theta(X_m) +
\zeta^\mu \partial_\mu \Theta(X_m)$
such that the gravity path integral cannot be simply disentangled
from the integrals over the monopole collective coordinates. The infinite dimensional field
space $g,\Theta$ has been mixed up with the $6$ dimensional monopole 
coordinate $X_m,\Xmb$ space. There is no generic way to choose a 
$6$ dimensional slice through the extended field space.
At high temperatures there was a well defined way to do this split, and we left the monopole integrals to the end. Here we are forced to do the integrals at the same time as the gravitational path integral.  These complications appear to be related to the usual issues of not having
local observables in quantum gravity.

At this stage we note that it seems hard to solve (\ref{monpos}) in the perturbative $\hat{q}_m/L$ 
expansion we outlined in (\ref{pertf}-\ref{pertf2}) since $\partial_\x \hat{\Th}$ is non-zero at zeroth order due to the
background electric field. 
Thus it seems that in order to satisfy (\ref{monpos}) one must work outside
the perturbative expansion - effectively canceling the zeroth order
term with the first order $(\hat{q}_m/L)$ interaction term.
It might be possible to do this if we move the monopole anti-monopole pair very close together where
the interaction potential is large: this is physically reasonable, essentially the Berry phase effectively binds monopole-antimonopole pairs together and prevents them from proliferating. However this is not the physics we are looking for; rather we seek a contribution to a holographic two-point function, which should depend on physics determined by monopoles separated by a large distance. 

\subsubsection{High momentum bulk to boundary propagator}

There is one further ingredient in our holographic computation: the bulk-to-boundary
propagator, which as we will see should have been included in the saddle.  In the regime of interest the momentum flowing in the propagator is very
large:
\be
p \sim 2 k_F = \frac{1}{16 \pi G_N} \hat{q}_m \hrho \,.
\ee
Thus this {\it propagator} will actually back-react on
gravity in a well defined way which we explore now. Indeed it
appears at exactly the same order as the Berry phase term and thus can 
resolve the issue of satisfying \eqref{monpos}. Following \cite{Louko:2000tp} we write
a ``massless worldline'' path integral representation for the bulk to boundary propagator from a point $y$ on the boundary to a bulk point $X_m$ as:
\be
\label{Kpath}
K^{(g)}(X_m;y) = \int_0^\infty \frac{dl}{l^2} \int [\mathcal{D} \bar{X}(s)] \exp\left(-  \int_0^{\,l} ds  g_{\mu \nu} (\bar{X}(s)) \frac
{d \bar{X}^\mu}{ds} \frac{d \bar{X}^\nu}{ds} \right)\,.
\ee
Here the integral is over all paths $\bar{X}(s)$ from the boundary to the specified point in the bulk. $s$ is a parameter along the worldline, and the boundary conditions at $s=0$ fix the path to the boundary at
\be
\bar{r}(0) = r_\Lambda \,, \qquad \bar{X}^a(0) = y^a\,,
\ee 
where $r_\Lambda$ is the UV cutoff. The boundary conditions at $s=l$ fix
the path to a point $X_m$ in the bulk $\bar{X}^\mu(l) = X^\mu_m = (r_m,y^a_m)
= (r_m,\tau_m,x_m)$. We have essentially fixed a gauge for the einbein leaving
an integral over $l$.

The utility of $K^{(g)}$ is that it allows us to define the bulk
to boundary propagator for a general metric $g$ (asymptotically AdS). 
This would be formally difficult in terms of solving the PDE wave equation on some
general metric $g$. Also since the momentum in the world line is necessarily large we expect the ``geodesic'' approximation to the path integral to be a good one. 

Because it is not clear what we might mean by
a ``massless geodesic in Euclidean space'' we consider a simple example which 
demonstrates that in general we must consider \emph{complex geodesics}.
These can be understood as complex saddles of \eqref{Kpath},
a prescription for which can be given in terms of the usual WKB approximation
to the wave equation \cite{Festuccia:2005pi}. 

We take the metric to be the charged BTZ metric.
The saddle equations for the path $\bar{X}(s)$ follow from the conserved charges associated
with translation invariance $E,P$ and the massless condition:
\be
\label{zerogeo}
E = 2 f(\bar{r}) \frac{d\bar{\tau}}{ds} \,,\qquad P = 2 \bar{r}^2 \frac{d \bar{x}}{d s}\,, \qquad
\frac{1}{f(\bar{r})} \left(\frac{d \bar{r}}{ds}\right)^2 + f(\bar{r}) \left( \frac{d\bar{\tau}}{ds} \right)^2 + \bar{r}^2 \left( \frac{d \bar{x}}{ds}\right)^2 = 0\,.
\ee
Note that the equation of motion for $l$ imposes the massless geodesic
constraint\footnote{The variation of the single number $l$ imposes the constraint at one point, but it is preserved by the dynamical equations of motion and so must now be satisfied everywhere.}.  

Now consider taking the Fourier transform of $K$ with respect to the boundary
coordinate:  
\be
\label{fpsad}
f_p(X_m)  = \int d^2 y\;e^{-i p_a(y^a - y_m^a) } K(X_m,y) \,.
\ee
Again performing this integral by extremizing with respect to $y^a$, we see that this fixes the charges in \eqref{zerogeo} to be purely imaginary $E = i \omega_E$ and $P = i k$. The appropriate geodesic thus follows an imaginary $\bar{x}(s), \bar{\tau}(s)$ trajectory. On this complex geodesic \eqref{fpsad} computes the WKB approximation to the massless wave equation
\cite{Festuccia:2005pi}.
For example if we set $\omega_E = 0$ the path \eqref{zerogeo} written in terms of $\bar{x}(\bar{r})$ is simply given by
\be
\frac{d \bar{x}}{d \bar{r}} =-  \frac{ i\, {\rm sgn}(k)}{\bar{r} \sqrt{f(\bar{r})}}\,,
\ee
where the sign is determined by the requirement that $(d\bar{r}/ds) < 0$.
Then \eqref{fpsad} becomes,
\be
f_p(X_m) \sim \exp\left( - |k| \int_{r_m}^\infty \frac{dr'}{r' \sqrt{f(r')}} \right)\,.
\ee
This is the WKB approximation to the massless wave equation in the
charged BTZ black hole at large $k$. For $r$ large it reduces to the
result we used previously in \eqref{f} for the bulk to boundary propagator, where
it is clear we should take the decaying WKB solution for regularity in the interior. 
The complex geodesic is telling us that the bulk to boundary propagator is highly suppressed
at high momentum as we move into the bulk.

Having built some confidence in $K^{(g)}$ we now include them in our path
integral \eqref{z11g}. This path integral will then directly computes the density density correlation
function,
\bea
\left<\delta \rho(y_1) \delta \rho(y_2) \right> \sim \frac{1}{Z^{(0,0)}} \int [\mathcal{D} g] [\mathcal{D} \hat{\Th}]
&& \hspace{-.4cm} \int d^3 X_m \sqrt{g(X_m)}  K_1^{(g)}(X_m,y_1) \hspace{-.1cm}   \int d^3 \Xmb \sqrt{g(\Xmb)} K_2^{(g)}(\Xmb,y_2)  \nonumber \\ 
&& \hspace{1cm} \times
\exp\left( - S[X_m,\Xmb] \right) \qquad +\,\, (m \leftrightarrow \bar{m}) \label{finsadg}  \,,
\eea
where we will not keep track of the overall normalization since we will only
compute the path integral in a saddle approximation anyhow. The notation
may get confusing, so we urge the reader to refer back to Fig.~\ref{fig:witten2}.
We use $I=1,2$ to label the two paths $\bar{X}_I$ and the two
boundary points $y_I$ where the currents are inserted . 
We use $i=m,\overline{m}$ to label the monopole/anti-monopole positions $X_i$.
For the particular saddle we consider here they are correlated as $m \leftrightarrow 1$,
$\overline{m} \leftrightarrow 2$. So for example below we always sum over these indices:
\be
\mathcal{S} = \{(i,I)\} = \{ (m,1), (\overline{m},2) \} \,.
\ee
The diagram Fig.~\ref{fig:manywittens}{\color{blue}B} has the opposite correlation.
The full set of equations that follow from varying everything
$\{g_{\mu\nu},\hat{\Th},\bar{X}_I,l_I,X_i\}$ are
\begin{gather}
\label{refdc}
R_{\mu \nu} - \frac{1}{2} g_{\mu\nu}\left( R + 2 \right) -   \ha T_{\mu\nu}^{\hat{\Th}}  =
  - \sum_{I=1,2} \hat{q}_m \int_0^{\hat{l}_I} d \hat{s}_I \frac{\delta_C^{(3)}( X - \bar{X}_I(\hat{s}_I))}{\sqrt{g}} \left[ \frac{d\bar{X}_I}{d\hat{s}_I} \right]_\mu
 \left[ \frac{d\bar{X}_I}{d\hat{s}_I} \right]_\nu\,, \\
 \label{theoma}
 \sqrt{g} \nabla^2 \hat{\Th} = \sum_{i=m,\overline{m}} \pm i \hat{q}_m  \delta_C^{(3)}(X-X_i) \,, \\
 \frac{d^2 \bar{X}^\mu_I}{d\hat{s}_I^2}+ \Gamma^\mu_{\alpha \beta} (\bar{X}_I)  \frac{d \bar{X}^\alpha_I}{d\hat{s}_I} \frac{d \bar{X}^\beta_I}{d\hat{s}_I} = 0 \,, \\ 
 g_{\mu\nu}(\bar{X}_I) \frac{d \bar{X}^\mu_I}{d\hat{s}_I}  \frac{d \bar{X}^\nu_I}{d\hat{s}_I}  = 0 \,, \\
 \label{pathbd}
 \partial_\mu \hat{\Th}(X_i) =   \pm 2 i g_{\mu\nu}(X_i) \left. \frac{d \bar{X}^\nu_I}{d\hat{s}_I} \right|_{\hat{s}_I = \hat{l}_I} 
\qquad (\mathrm{and}\,\, \bar{X}_I(\hat{l}_I) = X_i) \qquad (i,I) \in \mathcal{S}\,.
\end{gather}
In  \eqref{theoma}  and \eqref{pathbd} the $+$ refers to the coupling the monopole at $X_m$ and the $-$ to the antimonopole at $\Xmb$. It was also natural to rescale $(s_I,l_I) =
(16 \pi G_N/\hat{q}_m) (\hat{s}_I, \hat{l}_I)$ similar to the rescaling of $\Th \rightarrow \hat{\Th}$. 
The two paths are fixed to the boundary at
$\bar{r}_I(0) = r_\Lambda, \,\bar{y}^a_I(0) = y_I^a$.
 It can be checked that the full stress tensor coupling
to gravity is conserved when all the above equations are satisfied.

The equation \eqref{pathbd} is the endpoint equation of motion for the position
of the path, or really the equation of motion resulting from varying the
position of the monopoles. This is the generalization of \eqref{monpos} that 
we had trouble satisfying before. The missing momentum at the location of the monopole is now being supplied by the bulk-to-boundary propagator. One can heuristically think of the bulk to boundary
paths as strings attached to monopoles from the boundary that allow us to pull the monopoles
apart. As we shall see, this can be achieved only if we probe the monopoles at the correct (i.e. Friedel ``$2 k_F$'' wavevector)
momentum.

Finally some care must be taken to define the delta function $\delta_C^{(3)}$ in \eqref{refdc} 
since the path $\bar{X}_I(\hat{s_I})$ will generally follow a complex trajectory. The best
way to proceed is to solve Einstein's equation on complex coordinates
which pass through these $\bar{X}_I$ trajectories. It seems possible to understand this
in the perturbative expansion of $\hat{q}_m/L$ outlined above and explored
further below.
Some inspiration from the real time
formalism of thermal field theory (i.e. the Schwinger-Keldysh contour) will probably be necessary.

\subsubsection{Some aspects of the final saddle}

We can solve (\ref{refdc}-\ref{pathbd}) using the perturbative framework
outlined in Section \ref{pertframe}. In addition to the expansions
of the field $\hat{\Th}$ and $g$ given in (\ref{pertf}-\ref{pertf2}) we should also write:
\bea
\bar{X}_I(\hat{s}_I) &=& \bar{X}_I^0(\hat{s}_I) + \mathcal{O}(\hat{q}_m/L)\,,  \\
\hat{l}_I &=& \hat{l}_I^{\,0} + \mathcal{O}(\hat{q}_m/L) \,,  \\
X_i &=& X_i^0 + \mathcal{O} (\hat{q}_m/L)  \,,
\eea
where we will only need the zeroth order term and thus we
drop the $0$ superscript from now on. Then the procedure for constructing
the saddle is simple. We start with the zeroth order solution, which for $g,\hat{\Th}$ is simply
the charged BTZ black hole. The zeroth order geodesics are then the same
as discussed around \eqref{zerogeo},
where the rescaled conserved charges $\hat{E}_I,\hat{P}_I$ are fixed by the
monopole equation of motion \eqref{pathbd},
\be
\hat{E}_I \equiv 2 f(\bar{r}_I) \frac{d\bar{\tau}_I}{d\hat{s}_I} = 0 \,, \qquad
\hat{P}_I \equiv 2 \bar{r}_I^2 \frac{d\bar{x}_I}{d \hat{s}_I} =  \mp i \hat{\rho}
\qquad \rightarrow \qquad  \frac{d\bar{x}_I}{d\bar{r}_I} = \pm i \frac{1}{\bar{r}_I \sqrt{f(\bar{r}_I)}} \,.
\ee
Note that the momentum in the geodesic is fixed by the monopole equation of motion to be the Friedel wavevector. It
will be often convenient to parameterize the worldline
in terms of $\bar{r}_I$ instead of $\hat{l}_I$, it should be clear which one we are using.
The bulk positions of the monopole can now be found by solving for the the end point of the above paths:
\be
\bar{\tau}_I(r_i) \equiv \tau_i = \tau_I \,, \qquad \bar{x}_I(r_i) \equiv  x_i =x_I \mp i \int^\infty_{r_i} \frac{dr'}{r' \sqrt{f(r')}}  \qquad (i,I) \in \mathcal{S}\,,
\label{monpossolved}
\ee 
where $(\tau_I,x_I) = y^a_I$ are the positions of the boundary insertions.
The radial coordinate of the monopoles $\bar{r}_I(\hat{l}_I) \equiv r_i$ 
is fixed by the final monopole equation of motion,
the $r$ component of \eqref{pathbd}. 
\be
\label{comeb}
\frac{\hat{\rho}}{r_i \sqrt{f(r_i)}}  = \pm i \partial_r \hat{\Th}(r_i ,\tau_i,x_i)\,.
\ee
It should be clear that by symmetry the two $r_i$ are the same $r_m = \rmb$.
We will come back to this equation shortly. Finally to find $\hat{\Th}$ and $g$ to first
order we plug the zeroth order geodesic back into Einstein's equations and the dynamical equations for $\hat{\Th}$ and linearize. This will then provide sources for the equations that we discussed
in Appendix \ref{back} . Using the paths described above the sources simplify to the following:
\bea
\hspace{-1.1cm}
\label{linein}
\left( R_{\mu \nu} - \frac{1}{2} g_{\mu\nu}\left( R + 2 \right) -   \ha T_{\mu\nu}^{\hat{\Th}} 
\right)^1 &=& - \frac{\hat{q}_m \hat{\rho} }{2 \sqrt{f}} \theta(r - r_m)  \sum_{I=1,2}  n_\mu^I n_\nu^I\,   \delta(\tau -\tau_I)
\delta_C(x - \bar{x}_I(r) ), \\
\left( \sqrt{g} \nabla^2 \hat{\Th} \right)^1  &=& i \hat{q}_m  \delta( r- r_m)    \smmb \pm \delta (\tau - \tau_i)
\delta_C ( x - x_i) \,,
\label{madel}
\eea
where we have performed the integral over $\hat{s}$ by expressing it as an integral over $\bar{r}$ and $n^I_{\mu} = (dx \mp i \frac{\sqrt{f}}{r} dr)$ is a null vector pointed along the complex geodesic.
At this stage if we were to attempt to solve these equations we would need to confront the meaning of $\delta_C$. It can be understood in terms of a contour in complex $x$
space which passes through both monopole paths $\bar{x}_I(r)$ for $r>r_m$. For $r<r_m$ the
contour can lie on the real axis. Then these equations should be solved on this complex
contour. 

Fortunately for us, and perhaps unfortunately for the formalism we developed, much of
the above discussion is not important for the final result. The reason can be traced
to \eqref{comeb}, where the right hand side is actually zero to the order in the
$\hat{q}_m/L$ expansion we are working at. So it does not seem possible
to solve for $r_m$. Rather one can argue that $r_m = \infty$ is the solution - 
the monopoles live at the boundary. In order
to probe the monopoles -- i.e. in order to couple them consistently to gravity -- we had to include the bulk to boundary propagators
carrying a large amount of momentum. Since they carry large momentum the effective
tension in the paths of the bulk quanta is large and this pulls the paths towards the boundary.
This is because these propagators are exponentially suppressed
in $G_N$ upon moving in to the bulk  and so the saddle approximation
to \eqref{finsadg} will always want to concentrate the monopoles at the boundary. These are exactly the same considerations discussed when performing radial integrals in Section \ref{sec:highT} above. 

Let us now evaluate the answer for the current-current correlator in this framework. On-shell in \eqref{finsadg}, we see that the two monopoles in the bulk are actually located as close to the radial cutoff as they can be: furthermore, from \eqref{monpossolved} their positions in the field theory directions must coincide with those of the boundary theory operator insertions, i.e. $y_{m}^a \sim y_1^a, y_{\overline{m}} \sim y_2^a$. Thus we can simply evaluate the on-shell action with the monopoles held at this location. 

The calculation of the on-shell action itself was performed in the previous section. We can directly compute the correlator in position space, and we find (adding both diagram A and diagram B in Figure \ref{fig:manywittens} together)
\be
\langle \delta \rho(y_1)\delta \rho(y_2) \rangle \sim \cos\le(\rho q_m (x_1 - x_2)\ri) \le[(x_1 - x_2)^2 + v_s^2 (\tau_1 - \tau_2)^2\ri]^{-K},
\ee
where the prefactor cannot be reliably obtained in the saddle-point framework . This is equivalent to \eqref{finI}. Note that it may appear we have performed fewer Fourier transforms; however as the momentum-space expression is valid only in the neighbourhood of $k = \rho q_m$ in reality the information stored in these expressions is the same. 

We can go a little further than this by allowing
the first order correction to $\hat{\Th}$ to feature in \eqref{comeb}. By balancing these
two terms that appear at different orders in the $\hat{q}_m/L$ expansion
we will eventually find a true saddle for the monopole radial coordinates
at $r_m \sim \hat{\rho}/\hat{q}_m$. This is necessarily in the $AdS_3$ region
of the geometry. 

To see this we again argue that at large $r$ mixing with gravity is suppressed (by
factors of order $\hat{\rho}/r$.)  Additionally the worldline sources in \eqref{linein}
are suppressed. So we simply set the metric fluctuation at the boundary to zero $g_1 \approx 0$. 
The solution $\hat{\Th} = \hat{\Th}_0 + \hat{\theta}$ then follows from the matching procedure given in Section~\ref{friedmatch}. We write,
\be
\hat{\theta}(X) = i \hat{q}_m \left( G(X,X_m) - G(X,\Xmb) \right) \,,
\ee
where $G(X,X')$ 
was constructed in \eqref{neww}.
Plugging this answer
into \eqref{comeb} we find at large $r_m$,
\be
\frac{\hat{\rho}}{r_m^2}   = \left. \mp \hat{q}_m \partial_r \left( G(X,X_m) - G(X,\Xmb)  \right)
\right|_{X \rightarrow X_i} \approx -  \frac{\hat{q}_m}{2} \partial_{r_m} G(X_m,X_m)\,,
\ee
where we have used the fact that at large separation $G(X_m,\Xmb)$ is independent of $r$.
Note that the coincidence limit $G(X_m,X_m)$ gives the monopole self energy as discussed
in Appendix~\ref{sec:fug}.
We have not carefully constructed $G(X_m,X_m)$, however we can argue that
the dominant radial dependence comes from the $AdS_3$ answer since the self
energy is a local property which does not probe the full geometry. Reading
the $AdS_3$ answer from \eqref{fug_anal}, $2 \pi G(X_m,X_m) = \ln\ln(\bar{r}_\star/r_m)$, we find the position of the monopoles
in the saddle approximation is,
\be
\label{sadrm}
\frac{r_m}{ \ln(\bar{r}_\star/r_m) }= \frac{4 \pi \hat{\rho}}{\hat{q}_m} \,.
\ee
Surprisingly this saddle exists even in the high temperature calculation.
Examining the form factor $R(k)$ \eqref{rkint} we see there are two terms which depend on
$r_m$ in the exponential -  the bulk to boundary propagator and the monopole self energy.
These two terms work against each other, since the self energy wants to force the monopoles
into the bulk. The radial integral is then dominated in a saddle approximation 
at the same $r_m$ as in \eqref{sadrm}. 
Note that this contribution to the radial integrals is very different from the one we quoted in 
\eqref{rkfin} which was dominated by a UV divergence.
Roughly speaking in the $G_N \rightarrow 0$ limit there will be two contributions,
one from the UV divergence given in \eqref{rkfin} 
and one from the saddle at $r_m$ given in  \eqref{sadrm}. Which one wins
depends on the details of the UV completion. We did not see the UV divergent
piece in the approach of this section since the saddle approximation does not account
for contributions from the boundaries of field space (in this case $r_m \rightarrow \infty$). 
These do not minimize the action, but may still have the largest contribution.

The precise meaning of this particular saddle value of $r_m$ is not clear to us. 
Physically one would expect
the prefactor of the $2 k_F$ oscillations to be sensitive to the UV cutoff, however
here we are claiming a universal contribution in the $G_N \rightarrow 0$ limit.


\section{Conclusion} \label{sec:conc}

We now briefly summarize our results so far. We computed the contribution to the holographic density-density correlator from monopole events in the bulk on a charged black hole background, dual to a field theory state with a charge density $\rho$. Essentially we computed the Witten-esque diagram in Figure \ref{fig:witten1} by placing a monopole-antimonopole pair in the bulk and integrating them with the appropriate action cost over the whole spacetime. Independent of the details of the bulk geometry, we saw that each monopole event had a Berry phase \eqref{berry} associated with it arising from the coupling to the background electric field:
\be
S_{berry}[X_m] = -iq_m \rho \x_m,
\ee
where $\x_m$ is the spatial coordinate of the monopole in the direction parallel to the horizon. This oscillatory phase directly translates into a singularity in the density-density correlator in momentum space at the precise wavevector $k_{\star} = \rho q_m$. The precise form of the singularity depends on the monopole-antimonopole correlation in the bulk and thus on the geometry in question. We find at zero temperature:
\be
\langle \delta \rho(p) \delta \rho(-p) \rangle_{T = 0} \sim  \le(v_s^2(k - q_m\rho)^2 + \om_E^2\ri)^K +\; (k \to -k) \label{ffinal}
\ee
and at very high temperatures,
\be
\langle \delta \rho(p) \delta \rho(-p) \rangle_{T \gg \mu} \sim \frac{1}{(k- q_m \rho)^2 + (2\pi K^T)^2} + \; (k \to -k) \ . \label{ffinal2} 
\ee

Here the unwritten prefactors of these expressions are both exponentially small and non-universal in that they depend on the UV cutoff of the theory. $v_s$ is the speed of sound, and we have:
\be
K = \frac{1}{4\pi}\frac{q_m^2}{g_F^2}\frac{1}{v_s \log\le(\frac{r_{\star}}{2\hat{\rho}}\ri)}  \qquad K^T = \frac{1}{4\pi}\frac{q_m^2}{g_F^2}\frac{1}{\log\le(\frac{r_{\star}}{2\pi T}\ri)}\ . \label{Kdef}
\ee

\subsection{The existence of a Fermi surface}
We now finally turn to an interpretation of these formulas. We begin with the location of the singularity, for which we must first understand the meaning of $q_m$, the monopole magnetic charge. At the level of our current discussion it can be any magnetic charge that satisfies the Dirac quantization condition, $q_e q_m = 2\pi \mathbb{Z}$, where $q_e$ is the charge quantum associated with the compactness of the bulk $U(1)$ gauge group. Note that this $q_e$ has a direct {\it boundary} interpretation: it is the charge quantum appropriate to the {\it global} $U(1)$ symmetry of the boundary theory Hilbert space. Now it is commonly thought \cite{Polchinski:2003bq} that in theories of quantum gravity such as AdS/CFT, the spectrum of allowed bulk charges should be such that all Dirac conditions are {\it saturated}, i.e. the magnetic and electric charges should satisfy
\be
q_e q_m = 2\pi \ .
\ee 
We will not discuss in detail the arguments leading to this (see e.g. \cite{Banks:2010zn}), but we assume it to be true. 

Then the location of the Friedel oscillation singularity in $k$-space is
\be
k_{\star} = \frac{2\pi \rho}{ q_e} \ . 
\ee
This is precisely the value of $2k_F$ that we would find if we had constructed the entire charge density out of ordinary fermions with charge $q_e$. Luttinger's theorem -- interpreted here as a statement about a singularity in a gauge-invariant correlation function -- is {\it precisely satisfied}, despite the fact that there are no fermions in sight. There is something profound about the interplay between the field-theoretical Luttinger's theorem and the bulk Dirac quantization condition. It is also utterly clear that any deformation of the geometry will not change the location of the singularity: though $\rho$ can be tuned, the relation between $k_{\star}$ and $\rho$ is topological. 

There are several ways to interpret this result. One is tempted to associate this singularity with the presence of a dual Fermi surface. 
Note however that we do not have (nor do we need) access to fundamental fermions in the gravitational description: thus what we have actually shown is not the presence of a Fermi surface so much as the presence of structure in momentum space at a special momentum ``$2k_F$'' that is related to the charge density by Luttinger's theorem. This conclusion, while not previously demonstrated in holography, is not surprising. There exist elegant nonperturbative formulations of Luttinger's theorem that essentially prove this statement \cite{Oshikawa,Oshikawa2}, although there to make the discussion precise one needs to explicitly use the UV structure of the Hilbert space, i.e. the degrees of freedom living on an underlying lattice. In those works one then further argues that if the system is in a Fermi liquid phase one should identify ``$2k_F$'' with the location of the Fermi surface. In a more general context it seems possible that the desire to identify this momentum with ``the edge of occupied single-particle fermionic states'' is not a useful one. In some sense our holographic model provides us with a Fermi momentum with no reference needed to microscopic fermions. 

Of course in $1+1$ dimensions such a construction is not entirely unexpected. In standard field-theoretical treatments there is ultimately little difference between a density of fermions and bosons in one dimension: independent of the statistics of the microscopic degrees of freedom, they result in the same low-energy effective theory (that of the Luttinger liquid), and they both exhibit Friedel oscillations at the appropriate wavevector. Our holographic model appears to be a different way to arrive at a similar model, although the low-energy theory is different, as we elaborate on below. 

We should note from general principle of gauge-gravity duality we expect the microscopic Hilbert space should have some large number (``$N^2$'') of gauge-charged fermionic degrees of freedom. If each of these fermions individually formed a Fermi surface, then one would expect to see Friedel oscillations at each {\it individual} wavevector, which would be of $O(1)$ in the large $N$ limit. An example of this was given in the recent field theory model in one dimension \cite{Gopakumar:2012gd} of a ``strange metal'' -- a gauge theory of Dirac fermions in the adjoint representation. The model has a single $U(1)$ global symmetry, as in our setup, and a large-$N$ limit can be taken. They find Friedel oscillations which 
 preserve the original Fermi momentum of the underlying gauge-charged fermions,
 $2 k_F \sim \mathcal{O}(1)$. 
 This is not what we find: our wavevector scales like the total charge density and thus is of $O(N^2)$. Somehow in the gravitational description any putative individual gauge-variant Fermi surfaces are not visible: however the existence of a single special gauge-invariant momentum that counts the total charge is essentially guaranteed by the arguments of \cite{Oshikawa}, and reassuringly it appears in our holographic discussion. There is clearly more to be understood here. \footnote{
 One obvious
difference with our model is at low energy and high densities they recover a CFT.
We clearly do not have a CFT which would have separate left and right conserved currents. }

\subsection{Comparison to Luttinger liquid theory}
We turn now to more concrete considerations. In $1+1$ dimensions the Luttinger liquid theory provides a robust field-theoretical framework for strongly interacting physics at finite density. Starting from a microscopic fermionic Hamiltonian, one linearizes the fermion dispersion relation near each Fermi point, obtaining a separate left and right-moving sector that form a 2d CFT. The theory can be rewritten in terms of the sound mode, treating it as a free fundamental boson $\phi$, where interactions amongst the fermions modify the boson radius but are otherwise exactly marginal. The low-energy theory is thus characterized by two parameters: the speed of sound $v_s$ and the boson radius, parametrized by $K_{Lut}$. The density can be written in terms of this boson as (see e.g. \cite{Giamarchi}):
\be
\rho_{Lut}(y) = -\frac{1}{\pi}\nabla \phi(y) + \Lam\le( e^{2 i k_F\x} e^{-2i \phi(y)} + \mbox{h.c.}  + \cdots \ri). \label{lutans}
\ee
The first term is the gradient of the sound mode; the second results in Friedel oscillations, where $\Lam$ is a UV cutoff indicating the non-universal nature of the term, and the $\cdots$ indicate higher harmonics. 

It is interesting to compare our results to those of Luttinger liquid theory. The form of the Friedel oscillations in a Luttinger liquid are easily calculated from \eqref{lutans}. In the zero and high temperature limits they are of the same form as the our holographic results \eqref{ffinal} and \eqref{ffinal2}, with the substitution $K \to K_{Lut}$, $K^T \to v_s K_{Lut}$. In the Luttinger liquid these parameter relations are a consequence of $1+1$ dimensional conformal invariance, which completely fixes the form of the finite-temperature correlation. However the low-energy physics of our system is not a CFT, due to the marginally running coupling $\ka$ \eqref{rg1}. This running of the coupling explicitly manifests itself in the expressions for $K$ and $K^T$ \eqref{Kdef}, which contain extra scale-dependence relative to the Luttinger liquid case and can be thought of as measuring the value of $\ka$ at the scales corresponding to the density and the temperature respectively. We note that the precise dependence on the speed of sound is also different, again presumably due to its own logarithmic running \eqref{vs}.

It is interesting to note that the essential role played by the monopoles in our discussion is to attach the second term (or some caricature of it) in \eqref{lutans} to the holographic current operator. Indeed in field theory that term may be thought of as the contribution of vortices (i.e. instantons in 2d) in the field $\phi$ to its two-point function. These vortices are very small, probing the UV structure of the theory, and so their effect on low-energy physics can be completely understood in terms of a pointlike local CFT operator $e^{2i\phi}$. In holography the story is somewhat different. The monopoles in 3d play roughly the same role as the vortices in 2d, except that they can move freely in the holographic direction, corresponding to a field-theoretical instanton with no well-defined size and preventing an interpretation in terms of a local CFT operator. Despite this fact, the dominant overlap of the monopole field with the high-momentum probe is when the monopole is very close to the boundary, i.e. when the dual instanton is as small as it can be: this results in a non-universal coefficient for the Friedel oscillations similar to that in \eqref{lutans}. 

We note also that the zero-temperature compressibility of our model can be written in terms of $K$ and $v_s$ as \eqref{susc}
\be
\frac{d\rho}{d\mu}\bigg|_{T = 0} = \frac{q_e^2}{\pi} \frac{K}{v_s},
\ee
where as above we have assumed that $q_m$ saturates the Dirac condition. This relation is precisely that of Luttinger liquid theory (see e.g. \cite{Giamarchi}), with only a {\it single} species of fermion. It is fascinating that this {\it exact} relation still holds in our large $N^2$ theory, except that we now have a $K$ that is parametrically large, scaling with the number of charged degrees of freedom. In Luttinger liquid theory $K$ can be related to the strength of interactions of the microscopic degrees of freedom; it would be helpful to understand precisely what this means in our model.

We stress that while our model has some features in common with the Luttinger liquid, they are {\it not} equivalent: a Luttinger liquid is a pure CFT, and our marginally deformed theory must then be interpreted as describing a different phase of matter. Note for example that this system exhibits the familiarly distressing zero-temperature entropy associated with the near-horizon AdS$_2 \times \mathbb{R}$. This entropy does not play an important role in our analysis. Indeed there appears to be very little relation between the Fermi momentum structure described here, which is intimately associated with bulk gauge field dynamics, and the recently constructed geometries thought to describe hidden Fermi surfaces due to their entanglement structure \cite{Ogawa:2011bz,Shaghoulian:2011aa,HiddenFS}. It would be useful to understand the connection (if any) between these two approaches.

\subsection{Future directions}
There are several directions for future work, some of which we highlight here as we tie up some loose ends in our discussion. 

\ben
\item{\it Chern-Simons theory and UV completion}

As we have mentioned before, our usage of a Maxwell term in the bulk to describe a conserved current in two dimensions is somewhat nonstandard; as argued above this means that the theory is not conformal and will break down in the UV at a scale $r_{\Lam}$. One might instead be interested in studying a CFT, where a conserved current necessarily has separately conserved holomorphic and antiholomorphic parts. This is dual to the fact that it is represented in a 3d bulk by two gauge fields with Chern-Simons actions, i.e.
\be
S \sim k\int d^3 X\;(L \wedge dL - R \wedge dR)
\ee
In general $L$ and $R$ have Maxwell terms as well, but the dynamics is dominated by the Chern-Simons terms. Similar models were studied with an eye towards Luttinger liquids in \cite{Balasubramanian:2010sc, D'Hoker:2011xw}. 

We note here that the dynamics of monopoles is very different in Chern-Simons theories \cite{Pisarski:1986gr}. A Chern-Simons term associates flux with gauge charge, and thus a monopole creates charge as well as flux. A monopole and an anti-monopole are then necessarily connected by the worldline of a charged particle, which can be viewed as creating a tension-full string that confines them \cite{FradkinCSMon}. So monopoles do not proliferate freely in Chern-Simons theories, and the physics discussed in this paper will not obtain in quite the same manner. 

We see that this construction is not obviously related to the single propagating Maxwell gauge field that we studied. 
However if one considers this bulk theory in a Higgs phase where a diagonal (``axial'') combination of the gauge fields $R - L$ is Higgsed, then it turns out that the resulting theory has a propagating 3d gapless mode that can be described with a Maxwell action for the orthogonal (``vector'') current $L + R$ (see e.g. \cite{Mukhi:2011jp}). This Higgsing can be accomplished by introducing a scalar field $\Phi$ charged under the axial symmetry and explicitly sourcing it. The resulting physics is intricate, but if the scalar field flows to a constant in the infrared then the IR physics will likely be described by the Maxwell theory studied here. This provides a possible UV completion for this theory, and the interplay between monopoles and the Chern-Simons term is likely to lead to interesting physics in the holographic context. 

\item {\it Confinement in the bulk is dual to a charge gap on the boundary}

We turn now to a separate issue. In flat space (with no background electric field) in $(2+1)$ dimensions, the proliferation of monopole events in compact QED causes the gauge field to {\it confine} with a gap that is exponentially small in the monopole action \cite{Polyakov:1976fu}. It is natural to wonder what this means in the context of AdS/CFT: i.e. what {\it is} the dual of confinement of a gauge field in the bulk?\footnote{Some of the ideas presented in this section were arrived at during discussions with R. Loganayagam.} 

In such a phase bulk electric flux is confined into tight flux tubes, each of which has a finite tension per unit length. Now a state with a $U(1)$ charge in the field theory has electric flux at the boundary; however in the confined phase the only way to arrange this is to force one of these flux tubes to poke through the AdS boundary. The field-theoretical charge that we can get in this way is now quantized in units of the flux per tube $q_e$, and each tube extends into the bulk, meaning that each quantum of charge costs a great deal of energy. The previous sentence is of course precisely the description of a field theory with a {\it gap} to charged excitations. Confinement of a gauge symmetry in the bulk is thus dual to the formation of an {\it insulator} in the boundary; a convenient representation of this theory would be the Sine-Gordon model in the {\it bulk}:
\be
S_{SG} = \frac{g_F^2}{2}\int d^3 x \sqrt{g}\le( (\nabla \Th)^2 + \xi \cos(q_m \Th) \ri) \label{sg1}
\ee
Note $\Th$ has developed a potential, and flux tubes in the bulk correspond to kinks in $\Th$. 

Now we should ask whether or not such a phenomenon happens in our model. Note that in our model the background electric field results in the familiar Berry phase that causes the Friedel oscillations. On length scales longer than the inverse density the contribution of a monopole-antimonopole pair to the functional integral oscillates wildly in their relative distance. Integration over their relative coordinate would appear to wipe out the amplitude, essentially leaving behind only ``subtle'' effects (such as the Friedel oscillations discussed here). 

However, one could imagine canceling these Berry phases by considering a ground state which spontaneously breaks translation invariance. This would correspond to a lattice  of flux tubes carrying the electric flux from the black hole horizon to the boundary, where it is interpreted as a discrete set of localized charges. However, the spontaneous breaking of translational symmetry in a (1+1) dimensional field theory is not allowed by the Coleman-Mermin-Wagner theorem; thus at finite $N$ we do not expect this to be the ground state. More concretely, in the bulk there will always exist a gapless (in the field theory) Goldstone mode corresponding to fluctuations of this lattice, whose quantum effects are expected to destroy it. The resulting phase is what we have discussed in this paper: we do {\it not} believe that \eqref{sg1} is a useful starting point for understanding the finite-density state. The situation is rather different if we imagine {\it explicitly} breaking translation symmetry with a lattice: here one should be able to understand confinement (and hence holographic duals of insulators) at commensurate fillings. 

Note that most discussions of holographic ``insulators'' \cite{Nishioka:2009zj,Horowitz:2010jq,Balasubramanian:2010uw} describe a phase in which {\it all} excitations are gapped. Confining the bulk gauge $U(1)$ symmetry is different: only excitations that are charged under that $U(1)$ is gapped, indicating that the charge dynamics are playing an important role. This is a very desirable property for the holographic description of an insulating phase. A neutral sector can remain gapless and mediate long-range forces between the charged excitations. It would be interesting to understand such phases further. 

\item{\it Higher dimensions}

Finally, we turn to perhaps the most obvious extension of these results. As we have explained above, in $1+1$ dimensions the appearance of a Fermi momentum without access to fundamental fermions is -- while satisfying -- perhaps not tremendously surprising: a similar phenomenon happens in the traditional field-theoretical treatment of one-dimensional interacting {\it bosons}. In higher dimensions, of course, there is a true distinction between fermions and bosons, and the subsequent appearance of a Fermi surface from a higher dimensional charged black hole would be correspondingly more interesting. 

The natural question then is whether one can include fluctuations of the appropriate magnetically charged excitations in higher dimensions to show that {\it all} charged black holes contain momentum-space structure. This is of course considerably more difficult, as the dimensionality of the bulk excitation increases with dimension: in a $3+1$ dimensional bulk one needs to include monopole world-lines, i.e. a one-loop calculation involving quantum fluctuations of magnetically charged fields. While this may be technically challenging it does not seem impossible, and we hope to return to this and the other issues raised here in the future. 

The recent paper \cite{Hartnoll:2012ux} may provide useful guiding principles. In that work the electric flux through a bulk minimal surface is considered as an order parameter for fractionalization of charge. Indeed it seems that the $2 k_F$
observable we have constructed here is somewhat related to that order parameter. The monopoles play the role of the end points of the minimal surface (since the monopoles live in the UV of the geometry, this
analogy is more precise than it sounds). The electric flux integrated along
the surface gives
the same answer as the Berry phase term - 
in our construction the path taken between the monopoles is not important
because their are no explicit charges in the bulk (not behind the horizon.) 
 In the language of \cite{Hartnoll:2012ux} we are in a fully fractionalized phase. 
One big difference is that the Berry phase here
contributes an \emph{imaginary} part to the action, in contrast to the conjectured  \emph{real}
addition to the action of the minimal surface in \cite{Hartnoll:2012ux}. It would be interesting
to understand how the monopole story developed in this paper changes with
the addition of explicit bulk charges. We leave this to future work.

\een

We close on an optimistic note: the calculation outlined in this paper demonstrates that the class of charged black hole studied here behaves -- when probed in the right way -- as though it is made of charged excitations who are aware that quantum mechanics is encouraging them to form a Fermi surface. While there is clearly much more to be learned, we hope that the results presented here form another small step towards a useful application of holography to condensed matter physics. 

\vspace{1cm}

{\bf Acknowledgements}

\vspace{0.5cm}

We thank K.~Balasubramanian, F.~Franchini, S.~Hartnoll, G.~Horowitz, V.~Kumar, H.~Liu, R.~Loganayagam, D.~Marolf, J.~McGreevy, M.~Metlitski, J.~Polchinski, S.~Ryu, S.~Sachdev, B.~Swingle, and the Three Philosophers for valuable discussions, and J.~Santos for his enthusiastic encouragement.  This research was supported in part by the National Science Foundation under Grant No. PHY11-25915.
\begin{appendix}

\section{Magnetic monopoles, electric charges, and Berry phases} \label{sec:berry}

In this section we discuss some elementary aspects of the interaction of a magnetic monopole with a point electric charge in three spacetime dimensions. Some of our considerations can be found in a slightly different language in \cite{Metlitski:2007fu,MaxVBS}. In the bulk of the text we review the standard fact that the two actions
\be
S_A = \int d^3X \sqrt{g}\frac{1}{4 g_F^2} F^2\,, \qquad S_{\Th} = \int d^3 X \sqrt{g}\frac{g_F^2}{2}(\nabla \Th)^2,
\ee
describe equivalent theories, where the connection between the gauge field and scalar representations is \eqref{chiFrel}:
\be
\ep^{\mu\nu\rho}\p_{\mu} \Th = \frac{i}{g_F^2} F^{\nu\rho} \,.\label{chiFrel2} 
\ee
$S_A$ is the natural choice for including electric charges, where the coupling to a charge moving along a worldline $C$ is $S_A \to S_A + iq_e \int_C A$. Recall that this coupling tells us both how the fields respond to the charge (i.e. the Maxwell equations with sources), and how the charge responds to the field (i.e. the Aharanov-Bohm phase acquired along the trajectory $C$). Now consider working in flat space in polar coordinates,
\be
ds^2 = d\tau^2 + dr^2 + r^2 d\phi^2,
\ee
with a point charge at the origin $r = 0$. Then the electric field is $F_e^{r\tau} = i \frac{g_F^2 q_e}{2 \pi r}$. Turning to the scalar representation we see from \eqref{chiFrel2} that $\Th$ changes as we wind around the origin
\be
\Th_e(\tau,r,\phi) = \frac{\phi q_e}{2\pi} \,.
\ee 
For $\Th$ to be single-valued we should thus demand that it be a periodic variable $\Th \sim \Th + q_e$; this will make sense for all possible charges only if the gauge theory is compact with charge quantum $q_e$. 

Now consider adding a magnetic monopole source at a point $X_m$, which in the $S_{\Th}$ representation is the coupling $S_{\Th} \to S_{\Th} + i q_m \Th(X_m)$.  If we now {\it evaluate} this coupling on the field $\Th_e$ produced by an electric charge we find the relevant term in the action to be
\be
S_{berry} = iq_m \Th(X_m) = \frac{i q_m q_e}{2\pi} \phi_m \,.
\ee
This phase associated with a magnetic charge in an electric field is precisely analogous to the usual Aharaonov-Bohm phase acquired by an electric charge in a magnetic field. Each monopole event should be weighted by such a phase in the functional integral. Note this action appears to depend explicitly on the value of the periodic spatial coordinate $\th$, but it should of course be single-valued modulo $2\pi$; thus we conclude that we require
\be
q_m q_e  = 2\pi n \qquad n \in \mathbb{Z} \ .
\ee
This is the Dirac quantization condition. 

The form of this Berry phase means that in some loose sense the wavefunction of the system after the monopole-mediated transition depends on $\phi$, i.e. the monopole is creating angular momentum. We devote the rest of this Appendix to an understanding of this fact. 

First, we note that the canonical (Euclidean) stress tensor associated with $S_{\Th}$ is not conserved in the presence of sources: we have instead
\be
\nabla_{\mu}T^{\mu\nu}_{\Th} = g_F^2 \le(\nabla_{\nu}\Th \le[\nabla_{\mu},\nabla_{\nu}\ri]\Th + \nabla^2 \Th \nabla_{\nu} \Th \ri) \,.
\ee
Here the first term is nonzero whenever $\Th$ is multiple-valued, i.e. along the worldlines of electric charges. This is standard; demanding that this term (together with an appropriate contribution from the stress-energy of the particle itself) be conserved is equivalent to imposing the equation of motion of the charged particle in a background field. However, the second term is nonzero on the monopoles themselves, where we find
\be
\nabla_{\mu}T^{\mu\nu}_{\Th}(X) = i q_m \nabla_{\nu}\Th(X)\delta^{(3)}(X-X_m) = i \delta^{\nu\phi} \frac{q_m q_e}{2\pi}\delta^{(3)}(X-X_m)\,,
\ee
where the last equality is for the single charge/monopole configuration studied above. As claimed, angular momentum is being created at the location of the monopole. Note the requirement that this change in angular momentum be integer is another route to the Dirac condition \cite{Wilczek:1981du}. As described in Section~\ref{sec:monpos}, this fact can lead to considerable confusion if one wants to couple the system to gravity, which is somewhat unhappy with a non-conserved stress tensor. The resolution that we adopt is that one cannot simply fix the monopole position and study the monopole-charge system in isolation: to perform a fully consistent treatment one must supply the extra momentum from somewhere, which in our problem is via a bulk-to-boundary propagator from the AdS boundary\footnote{At some level the issues we are discussing -- concerning a perfectly reasonable-looking source who nevertheless violates a (contracted) Bianchi identity -- are in many ways similar to the issues that arise when trying to deal with a magnetic monopole while taking very seriously the identity of the electric gauge field $A$. One might then speculate that there is an alternative way to formulate the 3d gravity problem that elegantly deals with it, or at least some analog of a ``gravitational Dirac string'' that one could attach to the monopole, but we will resist the temptation to explore this here.}.

We conclude by noting that this creation of angular momentum seems more transparent if we consider the $U(1)$ gauge theory in a Higgs phase by condensing a scalar with charge $\frac{2\pi}{q_m}$, i.e. the value of the charge that saturates that Dirac condition. In this phase the monopole operator simply creates a massive state, a superconducting vortex with flux $q_m$. It is well-known that the composite of an electric charge and a vortex has a shifted angular momentum \cite{Wilczek:1981du} relative to the electric charge alone; thus the monopole operator must create it, just as the calculations above illustrate. 

\section{Boundary term in variational principle}
\label{sec:bdry}
Here we start with the bulk action in the form
\be
S = \frac{g_F^2}{2} \int d^3X \sqrt{g} \le(\nabla \th\ri)^2 + S_{bdy}
\ee
and construct $S_{bdy}$ to achieve a variational principle that is consistent with the boundary condition \eqref{bcsource}. We note first that the on-shell variation of the bulk part of the action is
\be
\delta S_{bulk} = -g_F^2 \int_{\p} d^2y\;\Pi\;\delta\th \,.
\ee 
However we would like the variation to be a function only of the source $H(y) = \Pi + \frac{1}{\ka}\nabla^2 \th$. To this end we note that if we take
\be
\label{sbdy}
S_{bdy} = g_F^2 \int_{\p} d^2 y \le(\Pi \th - \frac{1}{2\ka} (\nabla\th)^2\ri)
\ee
then the on-shell variation of the total action is
\be
\delta S = g_F^2 \int_{\p} d^2 y\;\th\le(\delta \Pi + \frac{1}{\ka}\nabla^2 \delta \th\ri) = g_F^2 \int_{\p} d^2 y\;\th \delta H,
\ee
as desired. Thus this is the correct form for $S_{bdy}$. Now the total contribution to the on-shell action (not just its variation) in the absence of any monopole sources can be computed to be 
\be
S_{\p} = \frac{g_F^2}{2}\int_{\p} d^2y\;\th(y) H(y)
\ee
If bulk monopoles are added then the action will receive extra contributions localized in the interior. In the main text this contribution is evaluated on a monopole-anti-monopole configuration. In this case we use \eqref{formsol} for the field $\th(x)$ as well as the relation between bulk-to-bulk and bulk-to-boundary propagators\footnote{This relation holds for any choice of boundary conditions.}: 
\be
G(r \to r_{\Lam},y_1;r_2,y_2) = K(r_2,y_2;y_1)
\ee
to find the boundary contribution to the monopole-anti-monopole action:
\be
S_{\p}[X_1,X_2] =  \ha \int_{\p} d^2y\;H(y)\le(iq_m (K(X_1;y) - K(X_2;y)) + g_F^2\int d^{2}y' K(r = r_{\Lam},y';y)H(y')\ri) \ . \label{bdycont}
\ee

\section{Computing monopole fugacities} \label{sec:fug}

In this appendix we discuss the computation of the monopole fugacity (i.e. the action cost of a monopole as a function of position) in curved space. We first discuss precisely what information is captured by the fugacity. Consider a monopole at $\Xm$ and an anti-monopole at $\Xmb$. It is shown in the text in \eqref{twobb} that the action of this configuration is formally
\be
S_{bb} = \frac{q_m^2}{2 g_F^2}(2G(X_m,\Xmb) - G(\Xm,\Xm) - G(\Xmb,\Xmb))
\ee
Here the first term is an interaction energy, and the second two terms appear to be singular self-energies. These terms contain nontrivial physics: for example, the charge-neutrality of this configuration guarantees that any IR divergences present in the interaction energy cancels against a corresponding divergence in the self-energies. It is in this sense that we can write an IR-cutoff independent answer for the propagator in expressions such as \eqref{monsep}. There is also UV information in these self-energies. In flat (or presumably any maximally symmetric) space it is understood that the singular part of this self-energy is entirely accounted for by a constant term $S_{c}$ that represents the action cost associated with the core of the monopole, and which can be computed given a UV completion of the theory. 

However if we break some of the spacetime symmetries then there is more information: while the UV divergence at the monopole location is the same (and thus there is still some action associated with the core) the interactions of the monopole field with the nontrivial geometry can result in a UV finite contribution that varies as we change the location $\Xm$ of the monopole.\footnote{It can be helpful to imagine the energy of a point charge near an infinite conducting plate.} This information is contained in the function $G(X,X_m)$, and we require a prescription to extract it, i.e. to perform the split:
\be
\frac{q_m^2}{2 g_F^2}\lim_{X \to X_m} G(X,X_m) \to -\le(S_{c} + S_{self}(\Xm)\ri) \equiv -S_m(r)
\ee
where $S_{int}(X_m)$ is the position-dependent contribution that we seek. This requires the subtraction of a divergence as $X \to X_m$, and it is important that this subtraction be unambiguous and introduce no extra spacetime dependence for the quantity $S_{int}(X_m)$ to have a meaning independent of the UV completion. 

To do this, we assume that we can incorporate the information regarding the core of the monopole in a function $F(X_m;X) \equiv f(d(X_m,X))$, where $d(X,X_m)$ is the {\it proper distance} from $X_m$ to $X$ and $f$ is thus a function of a single variable, where we take $f(y \to 0) \sim \frac{1}{y}$ to cancel the Coulomb divergence at the monopole core. The regulated interaction energy is then
\be
S_{self}(X_m) \equiv - \frac{q_m^2}{2 g_F^2} \lim_{X \to X_m} \le(G(X,X_m ) - f(d(X,X_m))\ri) \label{subt}
\ee
$f(y)$ is not unique; there are different choices which correspond to different choices for the internal monopole structure. However it is clear that only the finite part of $f(y \to 0)$ is important in determining $S_{int}(X_m)$, which is thus ambiguous only up to a constant that is independent of $X_m$. Thus we have separated the spacetime dependence from the UV structure of the monopole. By requiring that the subtraction depend only on the proper distance we are essentially asserting that the internal constituents of the monopole, whatever they may be, obey the equivalence principle and have no interactions with other fields. 

We now implement this prescription for the case of a monopole living in pure AdS$_3$. For pure AdS$_3$ the Green's function appropriate to the boundary conditions \eqref{bc1} can be calculated exactly in terms of Bessel functions to be 
\be
G(r,r';p) =   \frac{K_1\le(\frac{p}{r_<}\ri)\le(K_1\le(\frac{p}{r_>}\ri) - \log\le(\frac{p}{\bar{r}_{\star}}\ri)I_1\le(\frac{p}{r_>}\ri)\ri)}{r r' \log\le(\frac{p}{\bar{r}_{\star}}\ri)} \qquad p^2 = \om^2 + k^2 \qquad \bar{r}_{\star} = 2 r_{\star} e^{-\ga}
\ee
Note that $AdS_3$ is maximally symmetric, and thus the full radial dependence of the fugacity will come from the logarithmically running boundary conditions, i.e. the $r_{\star}$ dependence of the propagator. We first estimate this radial dependence. 

We begin by noting that the portion of this propagator that contains the term in $K_1 I_1$ does not contain any $r_{\star}$ dependence, and so appears conformal. One can check that this portion actually corresponds to the full propagator corresponding to conformal boundary conditions $\th(\infty) = 0$ at the AdS boundary. It thus respects the full symmetry of AdS space, and so once it is Fourier transformed to position space it is a function only of the proper distance between the two points. Following our discussion above, we see that this part of the propagator corresponds to the function $f$ defined in \eqref{subt}, and if we simply drop this part and take $r \to r'$ we will obtain the self-energy.  

\begin{figure}[h]
\begin{center}
\includegraphics[scale=0.8]{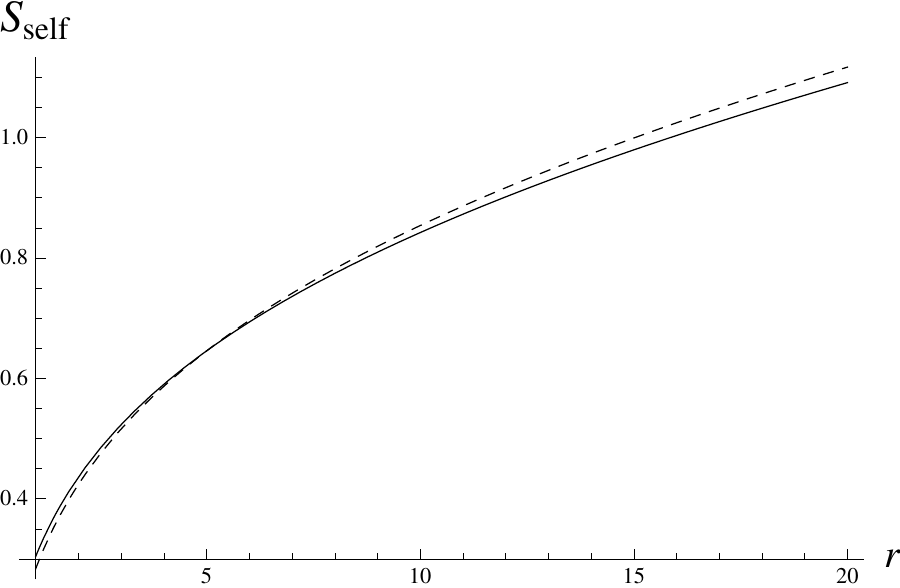}
\end{center}
\vskip -0.5cm
\caption{Numerical evaluation of monopole fugacity \eqref{fug1} (solid line) compared with analytic expression \eqref{fug_anal} (dotted line), with $\bar{r}_{\star} = 200$. As described above only the $r$-dependence can be computed within our framework, and thus an $r$-independent (and in fact IR-divergent) constant has been adjusted in the fit.}
\label{fig:fug}
\end{figure}

The self-energy is then
\be
S_{self}(r) = -\frac{  q_m^2}{4\pi g_F^2}\int_{0}^{\infty} p dp \frac{K_1\le(\frac{p}{r}\ri)^2}{r^2 \log \le(\frac{p}{\bar{r}_{\star}}\ri)} \label{fug1}
\ee
We will perform this integral in a rather crude manner which nevertheless (see Figure \ref{fig:fug}) captures the leading $r$-dependence. Note that at low $p$ the integrand is $\frac{1}{p \log\le(p \bar{r}_{\star}^{-1}\ri)}$ and at high $p$ the integrand is cut off by the exponential suppression of the Bessel function at $p \sim r$. We thus simply integrate the low $p$ asymptotic form from an IR cutoff at $L^{-1}$ to $p \sim r$ to find
\be
S_{self}(r) \sim -\frac{q_m^2}{4\pi g_F^2}\log\log\le(\frac{\bar{r}_{\star}}{r}\ri) \label{fug_anal}
\ee
In writing this down we have dropped an $r$-independent IR divergent term $\log \log \le(\frac{r_{\star}}{L}\ri)$ coming from the lower endpoint of the integration; as discussed above these divergences will cancel if we are looking at a charge-neutral configuration and so are not included in the definition of the fugacity. This double logarithm is slightly more palatable if one thinks of it as the logarithm of the proper distance from $r_{\star}$ to the monopole location. As $e^{-S_{self}}$ appears in the functional integral and we always have $r \ll r_\star$, it is clear that the monopoles want to be at small $r$, i.e. towards the interior. 

\section{Backreacted equations of motion}
\label{back}
Write the linearization of Einstein's and $\hat{\Theta}$ equations in the absence of sources
as:
\begin{gather}
\label{einlin}
E_{\mu\nu} = \left( -R_{\mu \nu}+\frac{1}{2} g_{\mu \nu} \left(R+2 - \frac{1}{2} (\nabla\hat\Theta)^2  \right) 
+ \frac{1}{2} \nabla_\mu \hat{\Theta} \nabla_\nu \hat{\Theta} \right)^1 = 0  \\
M = \left( \sqrt{g} \nabla^2 \hat{\Theta} \right)^1 = 0
\label{maxlin}
\end{gather}
where the superscript $1$ means linearize to first order.
The three constraints are:
\bea
\label{const1}
2 E_{rt} &=& i \omega_E \hxx'- i k \hxt' + \frac{f' r - 2f}{2f r} \left(2 ik \hxt  + i \omega_E \hxx \right)  \\
2 E_{rx} &=& \hrho \hat{\theta}' + i k \htt' -  \frac{r^2}{f} i \omega_E \hxt'
+ \frac{f' r - 2f}{2 f r} i k \htt \\
2E_{rr}  &=& -\frac{1}{r}\htt'- \frac{f'  }{2f}  \hxx'
+ \frac{k^2}{ fr^2}  \htt -\frac{2 k \omega_E }{f^2}  \hxt +\frac{\left(2 r^2 \omega_E^2+\hrho^2 f\right) }{2 f r^2} \hxx -\frac{i k \hrho  }{2f  r^2} \hat{\theta}
\label{const3}
\eea 
The four dynamical equations are:
\bea
\label{dyn1}
M &=& r f \hat{\theta}'' +\left(f' r+f \right) \hat{\theta}'  - \left(\frac{k^2}{r}+\frac{r \omega_E ^2}{f}\right) \hat{\theta} + \frac{i k \hrho  }{2 r} \htt  -\frac{i r \hrho  \omega_E  }{f} \hxt -\frac{i k \hrho  }{2 r} \hxx \\
2 E_{\tau\tau} &=&- f^2 \hxx''  -\frac{f f' r +4 f^2 }{2 r}\hxx' +  \frac{\hrho ^2 f }{2 r^2}\hxx -\frac{i k \hrho  f  }{r^2} \hat{\theta}   \\
2 E_{xx} &=& - r^2 f \htt''  -\frac{3}{2} r^2 f' \htt' -\frac{1}{2} \hrho ^2 \hxx+ i k \hrho  \hat{\theta} \\
2 E_{x \tau} &=& r^2 f \hxt''+  3 r f \hxt'  - \hrho ^2 \hxt+ i \hrho  \omega_E  \hat{\theta} 
\label{dyn4}
\eea
Collecting the first order fields in one big vector $X^I = \{ \hat{\theta},\htt, \hxx,\hxt,\hat{\theta}',\htt',\hxx',\hxt'\}$ with 8 entries we can write the above equations succinctly as a first order constrained system
\be
C_I^\mu X^I = 0 \qquad (X^I)' = \left(H_X\right)^I_J X^J 
\ee
where $\mu=(r,\tau,x)$ runs over the three constraints (\ref{const1}-\ref{const3}). Finally in this language we can write the equation for the gauge invariant variables $\sigma^\alpha$ given in \eqref{ginv} as,
\be
\sigma^\alpha = S^\alpha_I X^I
\ee
where $\alpha=1,2$.
The dynamical equations for $\sigma^\alpha$ in \eqref{eqsigma} derives from the following identity 
for the matrix $S$ (suppressing indices):
\be
S' =  H_\sigma S - S H_X  + \gamma C
\ee
where $\gamma_\mu^\alpha$ is a $2\times 3$ matrix and $H_\sigma$
is the $2\times 2$ matrix given on the right hand side of \eqref{eqsigma}. This equation and
in particular the matrices $H_\sigma$ and $\gamma$ can be derived by brute force.
Note that $\gamma$ is never needed so we do not write it here.

\section{Hydrodynamic limit}
\label{hydroapp}

We started studying the hydro limit in Section~\ref{hydlim} where a sound mode was found 
\eqref{hydjt}. Here we would like to go a little further and derive the width $\Sigma$ of the sound mode. We will only consider $T=0$ here, although it is certainly possible to include
a small $T$. 
Firstly we will need to go beyond the normalizable solution \eqref{hydsoln} of the hydro equation \eqref{hydeq} . So to begin with we find the second solution to \eqref{hydeq}. This can be done using standard tricks. The Wronskian of two solutions to \eqref{hydeq},  $(\sigma_1^A,\sigma_2^A)$
and $(\sigma_1^B,\sigma_2^B)$ has the following form
\be
\label{wronsk}
W = (\sigma_2^B \sigma_1^A - \sigma_1^B \sigma_2^A) = W_0 D_1(r)
\ee
where $W_0$ is a constant and $D_1(r) =  k^2 f' r + 2 r^2 \omega_E^2$. We can then use \eqref{wronsk} to integrate the unknown solution. The general
solution is
\be
\label{s1s2}
\begin{pmatrix} \sigma_1 \\ \sigma_2 \end{pmatrix}
=( a + b \, t(r) ) \begin{pmatrix} f \\  (r f' - 2 f)/\hat{\rho}^2 \end{pmatrix} 
+  b \frac{2D_1}{f \omega_E^2}   \begin{pmatrix} 0 \\1 \end{pmatrix}
\ee
where $a,b$ are integration constants and $t(r)$ is defined as,
\be
t'(r) = \frac{ 2 D_1 \hat{\rho}^2}{ r f^2 \omega_E^2}  \qquad
t(r \rightarrow r_h) = - \frac{4}{3} \left( \frac{r_h}{r-r_h} \right)^3 + \ldots + 0 \times ( r-r_h)^0 + \ldots
\ee
where the last condition fixes the integration constant in $t$. Once we have
$a,b$ we can simply read off the boundary quantity $\beta/\alpha$ which goes
into the final form of the current current correlator \eqref{Gmatch} 
\be
\label{beoval}
\beta/\alpha = \frac{2 (k^2 + \omega_E^2)}{\omega_E^2 (a/b + t(\infty) )} \,.
\ee

We now set out to find $b/a$. In order to find a width we need to effectively allow the sound mode to ``fall into the dissipative horizon.'' This happens at any non-zero $\omega_E$ since the hydro approximation to \eqref{eqsigma} necessarily breaks down at $(r-r_h) \sim |\omega_E|$. We should then match onto another solution close to the horizon. This procedure was used heavily in \cite{Faulkner:2009wj}
which we follow here closely. The ${\rm AdS_2} \times \mathbb{R}$ region makes this matching particularly nice
and has an interpretation in terms of RG flow of double trace deformations in the boundary field theory \cite{Heemskerk:2010hk,Faulkner:2010jy,Faulkner:2010tq}. 
 
The matching procedure requires splitting the radial direction up into inner
$0 < (r-r_h) \ll r_h$ (close to the Horizon or
AdS$_2 \times \mathbb{R}$ geometry) and outer regions $ (r-r_h) \gg |\omega_E|$
with solution given in \eqref{s1s2}. 
There is an overlap region where we match  if $|\omega_E| \ll r_h$. We also take $\omega_E \sim k$.
It is easiest to proceed using second
order equations for $\sigma^1$ which we can be derived from the first order equations 
\eqref{eqsigma}. In the inner region this equation becomes:
\be
\sigma_1''(\zeta) + \frac{2}{\zeta(8 \zeta^2 + 1)} \sigma_1'(\zeta) - \frac{8\zeta^2+1}{4\zeta^4} \sigma_1(\zeta) = 0
\ee
where $\zeta = (r-r_h)/\omega_E$. The regular solution at the horizon $r \rightarrow r_h^+$ is,
\be
\sigma_1(\zeta) = e^{-\frac{1}{2\zeta} {\rm sgn}(\omega_E)} (1 + 2  {\rm sgn} (\omega_E) \zeta + 4 \zeta^2 )
\ee
We can now expand $\sigma_1$ at large $\zeta$ and expand  \eqref{s1s2}
at small $r \rightarrow r_h$ and match.
\bea
\sigma_1(\zeta) &=& 4 \zeta^2 + \frac{1}{2} - \frac{1}{3} \zeta^{-1} {\rm sgn}(\omega_E)  + \mathcal{O}(\zeta^{-2})\\
\sigma_1(r) &=& 2 a (r-r_h)^2 \left( 1+ \mathcal{O}(r-r_h) \right)
- \frac{8 r_h^3 b }{3} \frac{1}{(r-r_h)} \left( 1+ \mathcal{O}(r-r_h) \right)
\eea
We can clearly match the two power laws $(r-r_h)^2$ and $(r-r_h)^{-1}$ (which
incidentally fix the conformal dimension of the appropriate operator in the $AdS_2$
theory.) There are various terms in the above expansions that do not match, these
were understood in \cite{Faulkner:2009wj} as coming from
perturbative corrections to both the inner and outer solutions. 
They are generally associated with certain analytic corrections to the real part of the sound mode dispersion relation, and will not effect the width \cite{Faulkner:2009wj}. So we can ignore
these and accurately read of the ratio,
\be
\label{bova}
b/a = \frac{1}{16} (\omega_E/r_h)^3  {\rm sgn}(\omega_E)
\ee
Putting everything together \eqref{bova}, \eqref{beoval} and \eqref{Gmatch} 
we find the width as defined in \eqref{hydjt} to be,
\be
\Sigma = \omega_E    {\rm sgn}(\omega_E)  \frac{(\omega_E^2 + k^2 )^2}{8 r_h^3 \ln(r_\star/r_h)}
\ee
Continuing this result to real frequencies we take only the $\omega_E>0$ section and
continue $\Sigma(\omega)$ in the upper half plane to get the retarded function
\be
\Sigma(\omega_E \rightarrow -  i \omega) \equiv \Sigma_R = - i \omega \frac{(-\omega^2 + k^2 )^2}{8 r_h^3 \ln(r_\star/r_h)}
\ee
For $r_\star > r_h$ it is clear that the sound mode appears in the lower half complex frequency plane as required by stability.

\end{appendix}

\end{document}